\documentclass[%
reprint,
superscriptaddress,
amsmath,amssymb,
prb
floatfix,
showpacs,
]{revtex4-2}

\usepackage{subcaption}
\usepackage{graphicx}
\usepackage{dcolumn}
\usepackage{bm}
\usepackage{hyperref}
\usepackage{ulem}

\usepackage[english]{babel}
\usepackage[utf8]{inputenc}
\usepackage{amssymb,amsfonts,amsmath,mathtools,mathrsfs}
\usepackage{relsize}
\usepackage{mdframed}
\usepackage[margin=0.6in]{geometry}
\usepackage{enumitem}
\usepackage{graphicx}
\usepackage{orcidlink}
\usepackage{float}
\usepackage{url,hyperref}
\usepackage[table,usenames,dvipsnames]{xcolor}
\usepackage{gensymb}
\usepackage{grffile}

\hypersetup{
	colorlinks=true,
	linkcolor=Blue,
	filecolor=magenta,      
	urlcolor=Blue,
	citecolor=Blue,
}
\usepackage[bottom]{footmisc}
\usepackage[capitalize]{cleveref}
\usepackage{textcomp}
\usepackage{braket} 

\setcitestyle{line}

\usepackage{amsmath,amssymb,wasysym}

\makeatletter
\def\@fnsymbol#1{\ensuremath{\ifcase#1\or *\or \dagger\or \ddagger\or
   \mathsection\or \mathparagraph\or \|\or **\or \dagger\dagger
   \or \ddagger\ddagger \else\@ctrerr\fi}}
    \makeatother

\begin{document}

\title{Room temperature intrinsic anomalous Hall effect in disordered half-metallic ferromagnetic quaternary Heusler alloy CoRuFeSi}

\author{Manikantha Panda}
\address{Department of Physics, National Institute of Technology Andhra Pradesh, Tadepalligudem 534101, India}

\author{Sonali S. Pradhan\orcidlink {0009-0003-7722-3720}}
\address{Department of Physics, Indian Institute of Technology Hyderabad, Kandi - 502285, Sangareddy, Telangana, India.}
\author{Prabuddha Kant Mishra}
\address{Current Affiliation: Institute of Low Temperature and Structure Research, Polish Academy of Sciences, Okólna 2, 50-422 Wrocław, Poland}
\author{Alapan Bera}
\address{Department of Physics, Indian Institute of Technology Kanpur, Kanpur 208016, India}
\author{Rosni Roy}
\address{UGC-DAE Consortium for Scientific Research, Kolkata Centre, Sector III, LB-8, Salt Lake, 700106, West Bengal, India}
\author{Rajib Mondal}
\address{UGC-DAE Consortium for Scientific Research, Kolkata Centre, Sector III, LB-8, Salt Lake, 700106, West Bengal, India}
\author{Soumik Mukhopadhyay}
\address{Department of Physics, Indian Institute of Technology Kanpur, Kanpur 208016, India}
\author{V. Kanchana\orcidlink {0000-0003-1575-9936}}
\email{kanchana@phy.iith.ac.in}
\address{Department of Physics, Indian Institute of Technology Hyderabad, Kandi - 502285, Sangareddy, Telangana, India.}

\author{Tapas Paramanik} \email[E-mail: ]{tapas.phys@nitandhra.ac.in}
\address{Department of Physics, National Institute of Technology Andhra Pradesh, Tadepalligudem 534101, India}

\begin{abstract}

Quaternary Heusler alloys offer a versatile platform for engineering magnetic and topological transport phenomena through chemical flexibility and tunable disorder. Here, we report a comprehensive experimental and theoretical investigation of the magnetic, magnetotransport, and anomalous Hall properties of the quaternary Heusler alloy CoRuFeSi. The compound crystallizes in the LiMgPdSn-type structure with significant Co-Ru antisite disorder and exhibits soft ferromagnetism with a saturation magnetization of $4.21~\mu_{\mathrm{B}}/\mathrm{f.u.}$ at low temperature and a Curie temperature well above room temperature. Hall measurements reveal a robust anomalous Hall effect persisting up to 300~K, with an anomalous Hall conductivity of $\sim 74$~S/cm that is nearly temperature independent. Scaling analysis demonstrates that the anomalous Hall response is dominated by the intrinsic Berry-curvature mechanism. First-principles calculations identify CoRuFeSi as a topologically nontrivial nodal-line semimetal in its ordered phase. Incorporation of experimentally relevant Co-Ru antisite disorder redistributes the Berry curvature and quantitatively reproduces the experimentally observed anomalous Hall conductivity, while preserving half-metallicity. These results establish CoRuFeSi as a disorder-tolerant half-metallic ferromagnet with a sizable intrinsic anomalous Hall effect at room temperature, highlighting its potential for spintronic and Hall-based device applications.

\end{abstract}



\maketitle

\section{Introduction}
The anomalous Hall effect (AHE), which results from broken time-reversal symmetry and spin-orbit coupling, is an additional contribution to the Hall response in magnetic materials that goes beyond the ordinary Hall effect. This contribution, which is proportional to the saturation magnetisation ($M_S$), acts as a sensitive probe of the underlying magnetic and electronic structure and produces a transverse voltage even in the absence of an externally applied magnetic field.
Particularly, the anomalous contribution to the Hall effect can result from both an intrinsic contribution and extrinsic scattering mechanisms (side-jump and skew scattering). 
In recent years, there has been considerable interest in the intrinsic AHE, as it is associated with electronic band topology \cite{SMIT195839,SMIT1955877,Berger1970sidejump,Karplus1954}.
In momentum space, Berry curvature acts as an effective magnetic field in systems with broken time-reversal symmetry and strong spin-orbit coupling (SOC). Large anomalous Hall conductivity (AHC) results from the sources and sinks of Berry curvature that occur at band crossings, such as Weyl or Dirac points close to the Fermi level \cite{nagaosa_review}. Consequently, AHE has become a sensitive diagnostic of topological electronic states, although the AHE is highly influenced by disorder and chemical ordering: whereas the intrinsic AHE is quite robust, structural flaws or antisite disorder can redistribute Berry curvature, shift or gap out band crossings, and inhibit or even reverse the AHC \cite{Shahi2022,Sinitsyn_disorder}.
In this domain, Heusler alloys provide exotic playground because of their strong SOC, broad chemical flexibility, and customizable electronic structure \cite{Heusler_review}, and shows a range of Hall responses, including topological Hall effects originating from noncoplanar spin textures, large AHEs associated with Berry curvature, and even quantum AHEs in carefully engineered thin films \cite{Co2MnGa_chang,Halfheusler_QAHE,Ajaya_2016}.

High-Curie-temperature soft ferromagnetic Heusler alloys are attractive for practical devices due to their low coercivity, minimal hysteresis losses, and stable room-temperature operation, which are crucial for magnetic sensors and spintronics based technologies \cite{Elphick31122021}.
In this context quaternary Heusler alloys (QHAs) (XX$'$YZ) are interesting possibilities for spintronic applications because they provide more chemical tunability, control over magnetic exchange, and electronic structure. Since partial atomic mixing and anti-site disorder are often unavoidable in these compounds, achieving half-metallic behavior against chemical disorder is challenging. Disorder tolerant half-metallicity ensures high spin polarization and reliable magnetotransport under realistic device conditions. Consequently, identifying QHAs that simultaneously exhibit robust half-metallicity and high-temperature soft ferromagnetism remains a central challenge for Heusler based spintronic materials.
In this context, the broad compositional flexibility of QHAs enables systematic tuning of symmetry-protected band crossings and band inversions through modifications in chemical ordering, magnetic configuration, and band filling. Furthermore, disorder itself can act as an additional tuning parameter, as antisite mixing or partial chemical disorder can modify the locations of Weyl points and symmetry-protected crossings, redistribute the Berry curvature, and broaden or shift inverted bands. Depending on the Berry curvature near the Fermi level, such disorder-induced modifications may either enhance or suppress the intrinsic Hall response \cite{Disordereffect_heusler}.

A prospective member of the QHA family, CoRuFeSi has recently been experimentally reported as a soft ferromagnetic half-metallic alloy \cite{BAINSLA2015631}, in which the half-metallicity and nearly 100 \% spin polarization are robust against antisite disorder \cite{kundu2017new,SEEMA2019106478}. This compound is also reported to possess the highest Curie temperature (867 K) among the QHA family \cite{BAINSLA2015631}.
In this study, we report the anomalous Hall effect and examine the impact of structural disorder on the AHE in CoRuFeSi. Through a comprehensive analysis of its magnetotransport properties, we focus on how disorder influences the anomalous Hall response. By combining experimental measurements with first-principles calculations, we elucidate how disorder modifies the electronic structure and redistributes the Berry curvature, thereby governing the anomalous Hall conductivity in this QHA.

\section{EXPERIMENTAL DETAILS}

Polycrystalline samples of CoRuFeSi were synthesized via a conventional arc melting technique under a high-purity argon atmosphere. High-purity elemental precursors (purity $> 99.9\%$) in stoichiometric proportions were used in the form of solid pieces. The obtained ingot was remelted five times, with intermittent flipping after each melting to ensure
compositional uniformity. 
Further, the as-cast ingot was enclosed in an evacuated quartz tube and annealed for five days at 800 °C, and then it was quickly quenched in freezing water. There was no discernible weight loss observed during the synthesis procedure.
A single-phase formation and crystal structure features were revealed by Rietveld refinement of room-temperature powder x-ray diffraction (XRD) data using the FULLPROF suite \cite{Fullprof}.
 With applied magnetic fields ranging from $\pm 7$~T, magnetic measurements were performed using a vibrating sample magnetometer (VSM), Quantum Design. From the bulk interior of the annealed ingot, a rectangular parallelepiped specimen with dimensions of 3.84$\times$1.9$\times$0.5 $mm^3$ was sectioned for magneto-transport measurements. Using a custom-built setup with an AC transport option, electrical transport measurements were performed using a standard six-probe approach with applied magnetic fields of 0 and 7 T spanning the temperature range of 2 to 300 K.

\section{Computational Details}
Geometrical optimization was performed using the Vienna ab initio simulation package (VASP) \cite{kresse1993ab, perdew1996generalized} within the density functional theory framework. The Perdew-Burke-Ernzerhof (PBE) exchange-correlation functional \cite{perdew1996generalized} within the generalized gradient approximation was employed. A plane wave energy cutoff of 600 eV was consistently applied throughout all calculations, with an energy convergence criterion set at \(10^{-8}\) eV. A $16\times16\times16$ k-point mesh was employed to sample the irreducible Brillouin zone (BZ) using the Monkhorst--Pack scheme \cite{monkhorst1976special}. To investigate the topological properties, a tight-binding Hamiltonian was constructed using maximally localized Wannier functions. The topological features were then analyzed using the iterative Green’s function method, as implemented in the \textsc{WannierTools} package \cite{mostofi2008wannier90, wu2018wanniertools}.

\section{RESULTS and DISCUSSIONS}

\subsection{Structure and magnetization}
\begin{figure}
    \centering
    \includegraphics[width=1\linewidth]{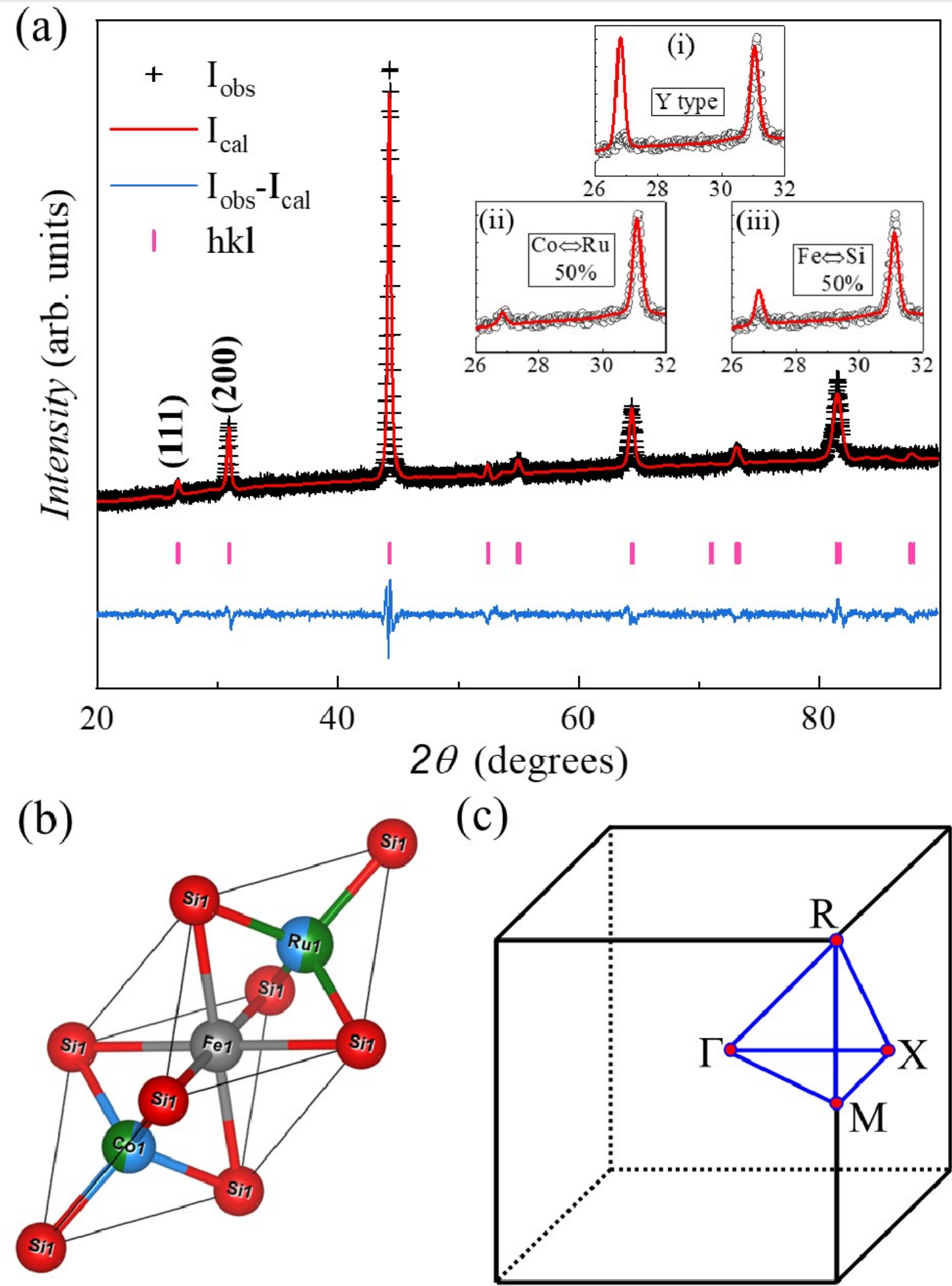}
    \caption{ Room temperature powder x-ray diffraction data of CoRuFeSi refined with Rietveld refinement. Vertical bars indicate the Bragg's reflection (pink), and the blue line indicates the difference between the observed data and the calculated data.  The insets show a zoomed view near (111) and (200) superlattice reflection peaks for different kinds of orderings. (i) XRD pattern with Y-type ordering, (ii) XRD pattern with 50$\%$ swap between the Co and Ru atom, and (iii) XRD pattern with 50$\%$ swap between the Fe and Si atom.  Refinement considering 50$\%$  disorder between octahedral site atoms Fe and Si in configuration (i). (b) Primitive unit cell corresponding to the fitted structure. (c) The bulk cubic Brillouin zone.} 
    \label{fig1}
\end{figure}


Powder x-ray diffraction (PXRD) studies were performed to ascertain the crystal structure of the equiatomic quaternary Heusler alloy (EQHA) CoRuFeSi, data shown in the \hyperref[fig1]{Fig. 1}. The experimental pattern shows a comparatively low intensity of the (111) superlattice reflection, with respect to (200) peak, suggesting the existence of antisite disorder in the system. Further, FULLPROF refinement was performed to determine the favorable ordered lattice.
CoRuFeSi crystallizes in the EQHA prototype, the LiMgPdSn-type structure (Space group: $F\bar{4}3m$, No. 216). This structure has four interpenetrating FCC sublattices along the body diagonal, formed by atoms occupying the Wyckoff sites 4a (0, 0, 0), 4b (0.5, 0.5, 0.5), 4c (0.25, 0.25, 0.25), and 4d (0.75, 0.75, 0.75). Three different ordered atomic configurations for the remaining three elements become apparent when one site is determined. Si was fixed at the site in consideration of the main group element's empirical propensity to occupy the 4d position \cite{ZHANG2023111541}. 
Depending on the atomic site preferences of $X$, $X'$, and $Y$, three ordered configurations YI, YII, and YIII are possible, with $X'$, $Y$, and $X$ occupying the $4c$ (0.25, 0.25, 0.25) site, respectively. 
From the refinement, the YIII structure provides the best fit, with a small discrepancy observed in the (111) and (200) superlattice reflections [see inset (i) of Fig.~\ref{fig1}(a)].
This discrepancy implies that the real structure is not perfectly ordered, prompting us to examine several disorder models. Previous studies on CoRuMnSi have shown a strong propensity for antisite disorder between Co and Ru \cite{VENKATESWARA2020166536}. In the present case, the agreement between the calculated and experimentally observed intensities of the superlattice peaks is significantly improved by introducing a 50\% site exchange between Co and Ru. In contrast, a 50\% Fe--Si antisite disorder fails to reproduce the observed diffraction pattern, likely due to the substantial difference in atomic radii and scattering factors between Fe and Si, which renders such disorder energetically unfavorable. Consequently, we infer that CoRuFeSi crystallizes in the LiMgPdSn-type quaternary Heusler structure with significant Co--Ru sublattice disorder, as shown in Fig.~\hyperref[fig1]{1(a)}.

Energy-dispersive x-ray analysis (EDX) was performed to verify the elemental stoichiometry, and the measured composition aligns well with the expected values within the permissible experimental error, as shown in \textcolor{blue}{Fig.S3}.

\begin{figure}
\begin{center}
    \includegraphics[width=1\linewidth]{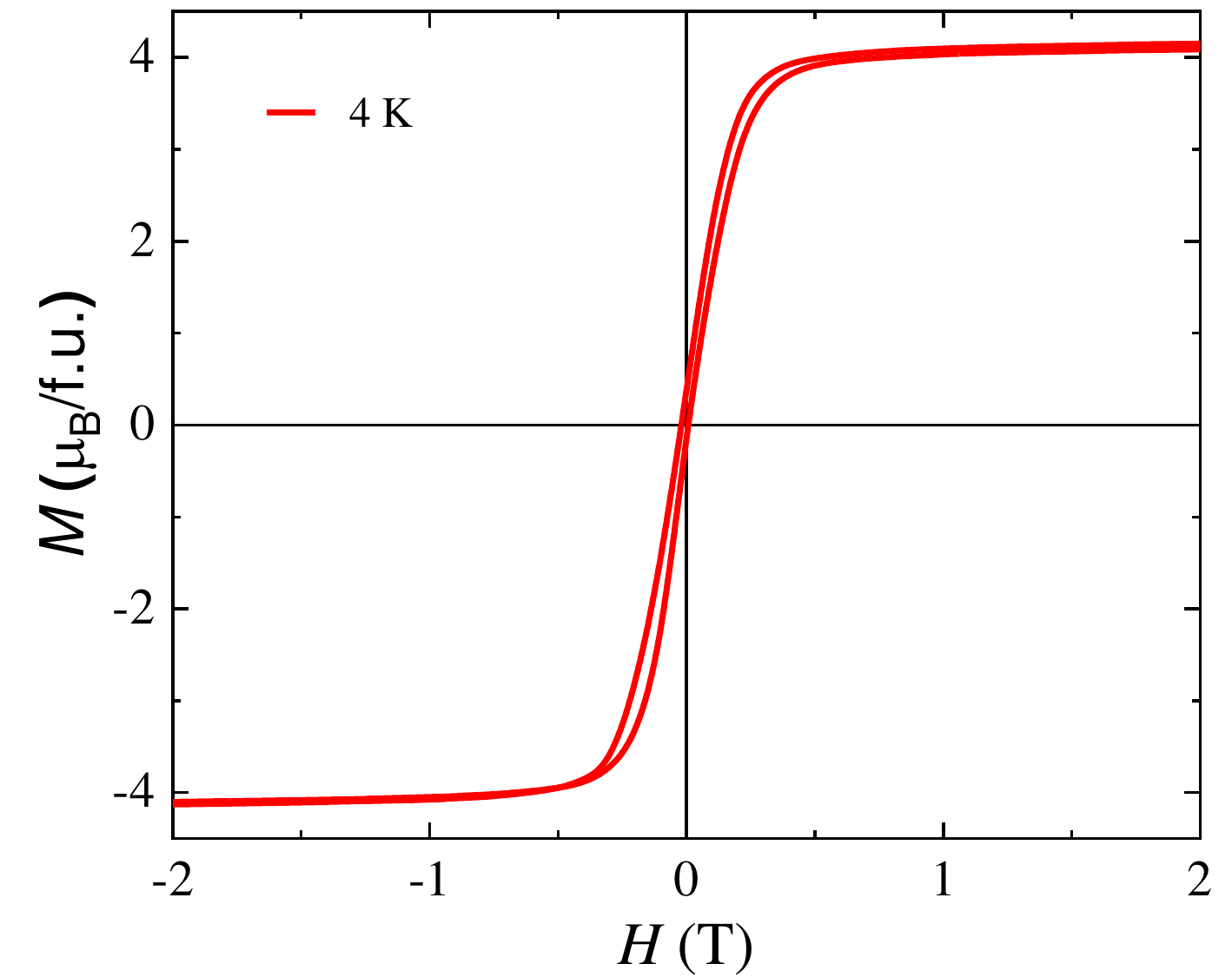}
    \caption{Isothermal magnetization $M(H)$ of CoRuFeSi measured at 4 K, showing soft-ferromagnetic behavior.}
    \label{fig2}
\end{center}
\end{figure}

The isothermal magnetization \(M(H)\) measured at 4~K is shown in Fig.~\ref{fig2}. The magnetization increases sharply with applied field and reaches saturation above 10~kOe, attaining a value of $4.21~\mu_{\mathrm{B}}/\mathrm{f.u.}$ at 4~K. The low coercive field observed in the hysteresis loop confirms the soft-ferromagnetic nature of CoRuFeSi, which persists up to room temperature, consistent with earlier reports \cite{BAINSLA2015631}.
Such soft ferromagnets are often proposed as promising platforms for hosting nontrivial topological electronic states \cite{Swekis2021,nayak2017magnetic}.

\begin{figure}
\begin{center}
    \includegraphics[width=1\linewidth]{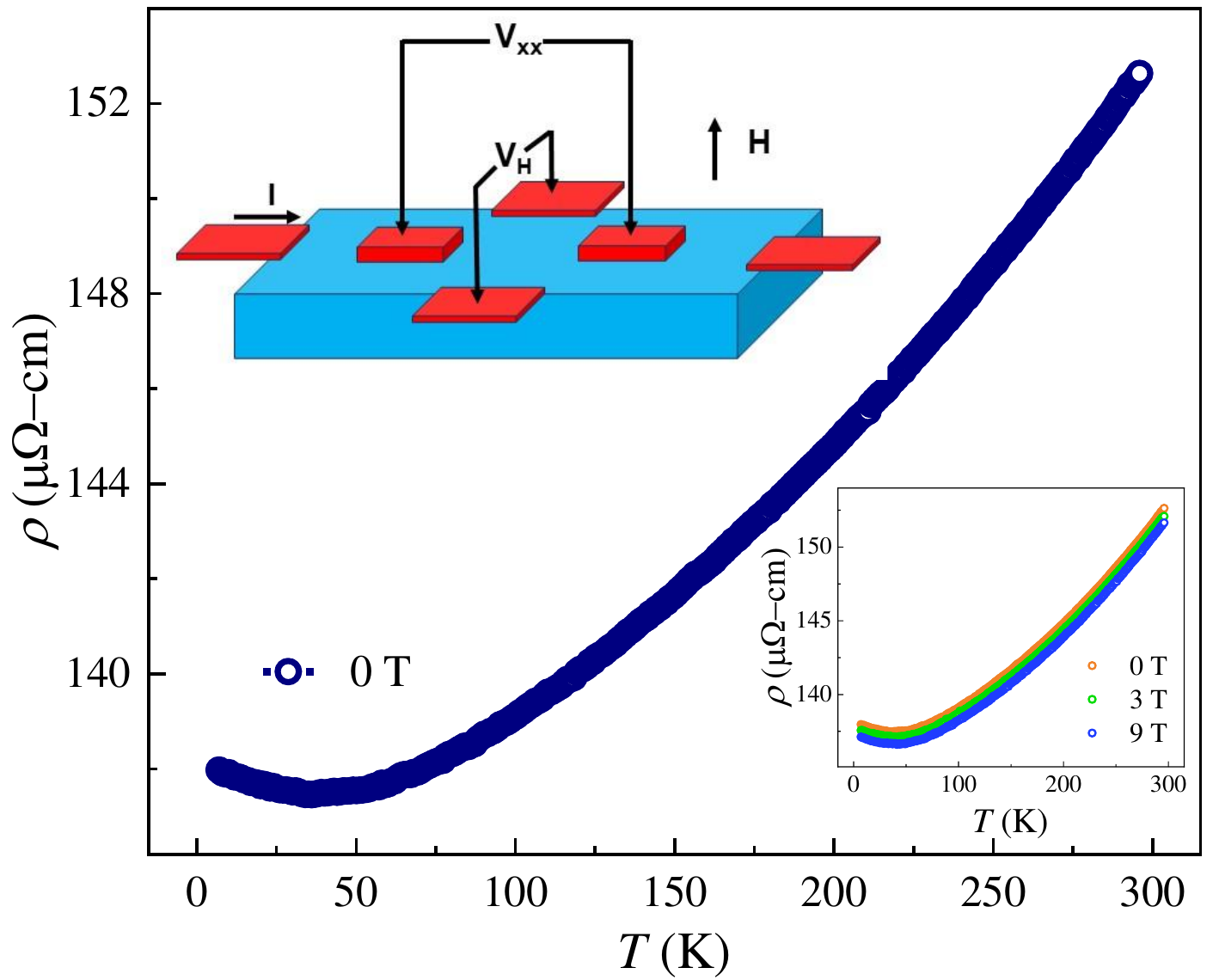}
    \caption{Temperature dependent longitudinal resistivity ($\rho_{xx}(T)$) with a schematic drawing of the technique of measurement used for Hall
voltage ($V_H$ ) and longitudinal voltage ($V_{xx}$ ) measurements. $\rho_{xx}(T)$. Inset shows the variation of resistivity under magnetic fields of 0 T, 3 T, and 9 T.}
    \label{fig3}
\end{center}
\end{figure}
\subsection{Electrical transport properties}
In this section, we studied the transport and magnetotransport behaviour, including the Hall measurement.
The temperature-dependent longitudinal resistivity ($\rho_{xx}(T)$) was measured both in the presence and absence of a field. The data is shown in \hyperref[fig3]{Fig. 3}. The metallic nature is obtained, as $\rho_{xx}(T)$ decreases gradually with lowering temperature. 
The residual resistivity ratio (RRR = $\rho_{300 K}/\rho_{2 K}$) is found to be 1.11, which signifies the presence of disorder in the compound. 
The high-temperature $\rho_{xx}(T)$ shows metallic behavior dominated by electron–phonon scattering, with a negligible electron–magnon contribution, consistent with half-metallic ferromagnetism \cite{VENKATESWARA2020166536}. 
On further lowering the temperature ($T<50~$K), $\rho_{xx}(T)$ exhibits a field-independent upturn, indicating dominant electron–electron interaction effects, likely associated with residual disorder as observed in related Heusler alloys \cite{Ti2FeAl_koushik}. A detailed  explanation of the $\rho_{xx}(T)$ behavior observed in this compound is described in the \textcolor{blue}{supplementary material}.

\begin{figure}
\begin{center}
    \includegraphics[width=1\linewidth]{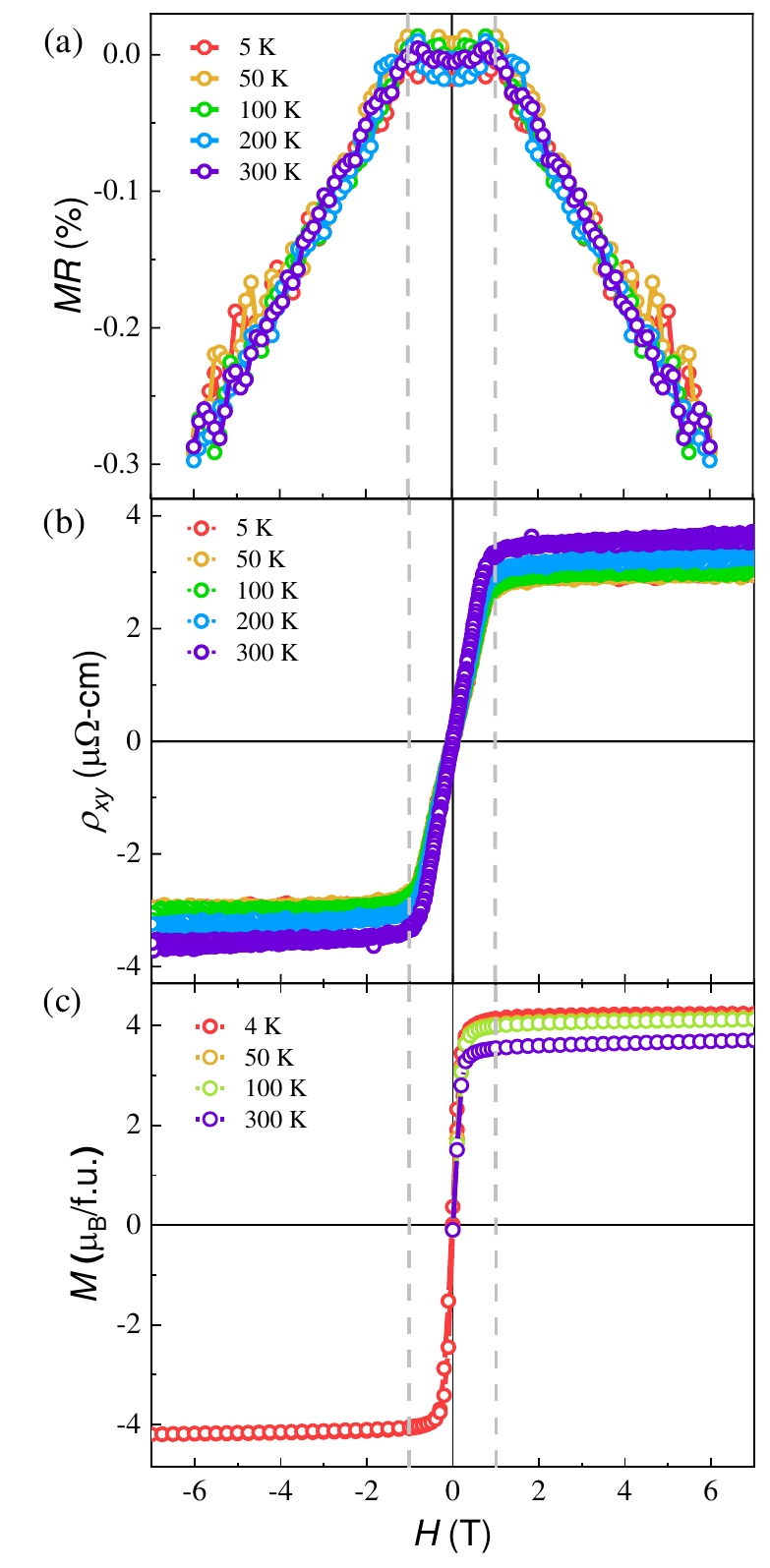}
    \caption{(a) and (b) Magnetoresistance (MR) vs H and Hall resistivity ($\rho_{xy}$) vs H at different temperatures, respectively. (c) Isothermal magnetization taken at 4, 50, 100 and 300 K for CoRuFeSi.}
    \label{fig4}
\end{center}
\end{figure}


\begin{figure*}
\begin{center}
    \includegraphics[width=1\linewidth]{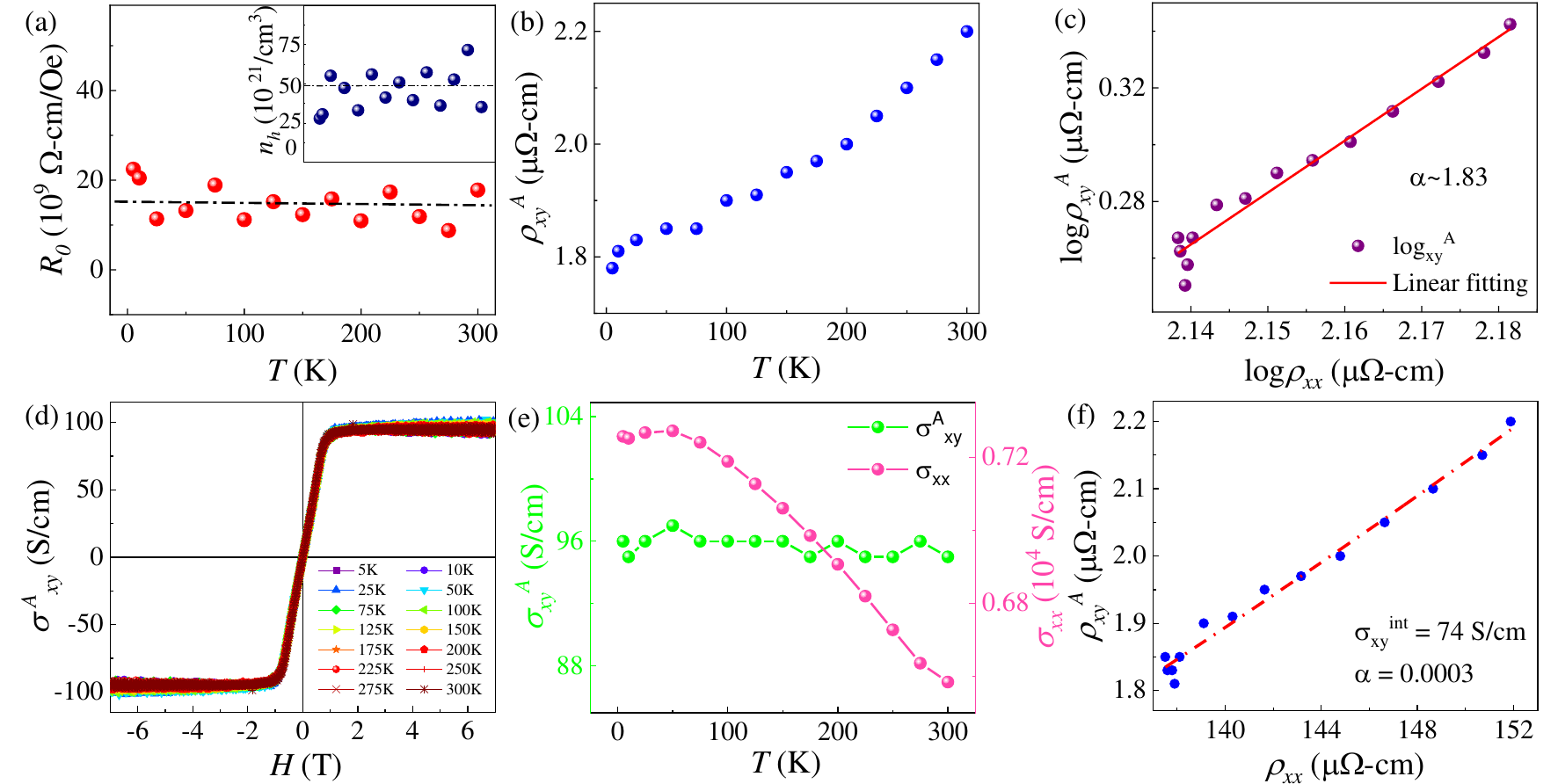}
    \caption{(a) Temperature dependence of the ordinary Hall coefficient $R_0$ (inset shows the corresponding temperature variation of the carrier density). (b) Temperature dependence of the anomalous Hall resistivity $\rho_{xy}^{A}$. (c) log--log plot of $\rho_{xy}^{A}$ vs $\rho_{xx}$, where the red solid line represents a linear fit to the data. (d) Anomalous Hall conductivity (AHC) as a function of magnetic field for different temperatures. (e) Temperature dependence of the AHC $\sigma_{xy}$ and longitudinal conductivity ($\sigma_{xx}$), and (f) $\rho_{xy}^{A}$ vs $\rho_{xx}$ with fitting to Eq.~(3), shown by the red dashed line.}

    \label{fig5}
\end{center}
\end{figure*}

Next, the magnetoresistance was measured as a function of magnetic field in the range of $\pm 6$~T, as shown in Fig.~4(a). The MR remains negligible up to an applied field of 1~T and then exhibits a nearly linear negative decrease for $H > 1$~T, reaching a value of about $-0.3\%$ at 6~T. Notably, the MR curves measured at different temperatures largely overlap and persist up to room temperature, which is consistent with the field-dependent longitudinal resistivity, $\rho_{xx}(H)$.
The most intriguing finding of this work is shown in \hyperref[fig4]{Fig. 4(b)}, which shows representative data for the Hall configuration as a function of the field. The linear field dependence is observed after the initial jump for a low applied field of 1 T. The jump is an anomalous contribution, which is considerable even at room temperature. 

A clear correspondence is observed in different measurements performed under applied magnetic fields of \(\pm 7~\mathrm{T}\) over a wide temperature range from 5~K to 300~K. The dotted line parallel to the \(y\)-axis at \(H = 1~\mathrm{T}\), extending throughout Fig.~\hyperref[fig4]{4}, highlights the correlations among the MR, $\rho_{xy}$ and M with respect to field. 
At this point, it is instructive to observe that the field evolution of the magnetization is closely related to the magnetoresistance at a given temperature. Because the spins are not perfectly aligned in the low-field region ($H < 1$~T), there is a modest positive magnetoresistance and increased spin-disorder dispersion. The spins align ferromagnetically when the field is increased beyond 1 T, which causes a minor change in magnetoresistance. Again, within the same field range, the Hall resistivity $\rho_{xy}(H)$ shows a significant increase that closely resembles the behavior of $M(H)$. From our experimental investigation, we establish that \(\mathrm{CoFeRuSi}\) exhibits correlated electronic behavior, reflected in the intertwined nature of its transport and magnetic properties. 



\subsection{Anomalous Hall effect}

To study the nature and origin of the AHE, we have studied the field dependence of Hall resistivity in detail, which is presented below. Fig. \hyperref[fig4]{4(b)} depicts the field dependent Hall resistivity ($\rho_{xy}$ vs $H$) data for CoRuFeSi compound. 
In general, total Hall resistivity can be written by the following equation;
\begin{equation}
    \rho_{xy}= \rho_{xy}^o+\rho_{xy}^A = R_0H+R_SM_S
\end{equation}
where the first term is due to the ordinary Hall effect, which is normally Lorentz force driven, and the second term is the anomalous contribution to the Hall effect ($\rho_{xy}^A$). In the above equation, $R_0$ is the ordinary Hall coefficient, $R_S$ is the anomalous Hall coefficient, and $M_S$ is the saturation magnetization. Initially, to calculate the anomalous Hall response, the ordinary Hall response is eliminated by linearly fitting the higher field data ($>$ 4 T), where the slope and intercept will give the value for $R_0$ and $\rho_{xy}^A$. The variation of $R_0$ with temperature is depicted in Fig. \hyperref[fig5]{5(a)}. The positive sign indicates that hole-type charge carriers primarily control the transport behavior of CoRuFeSi over the whole temperature range. The relation $R_0 = 1/ne$ can be used to determine the carrier density $n$ and the carrier type. The corresponding carrier density has been estimated based on the values of $R_0$ obtained and is shown in the inset of Fig. \hyperref[fig5]{5(a)}. The estimated carrier density at 5 K is 2.2$\times$10$^{22}$ per cm$^3$, which is in the good metal regime, and the corresponding carrier mobility ($\mu_h$) is found to be $\approx 2.1~\mathrm{cm^2\,V^{-1}\,s^{-1}}$. The obtained value of carrier density is reported for several other alloys, such as Mn$_3$Sn \cite{liu2018giant}, which exhibits the AHE. 

In \hyperref[fig5]{Fig. 5(b)}, the variation of $\rho_{xy}^A$ with respect to temperature is illustrated from 2 to 300 K, showing the monotonically increasing behavior of $\rho_{xy}^A$.  
\begin{figure}
\centering
    \includegraphics[width=1\linewidth]{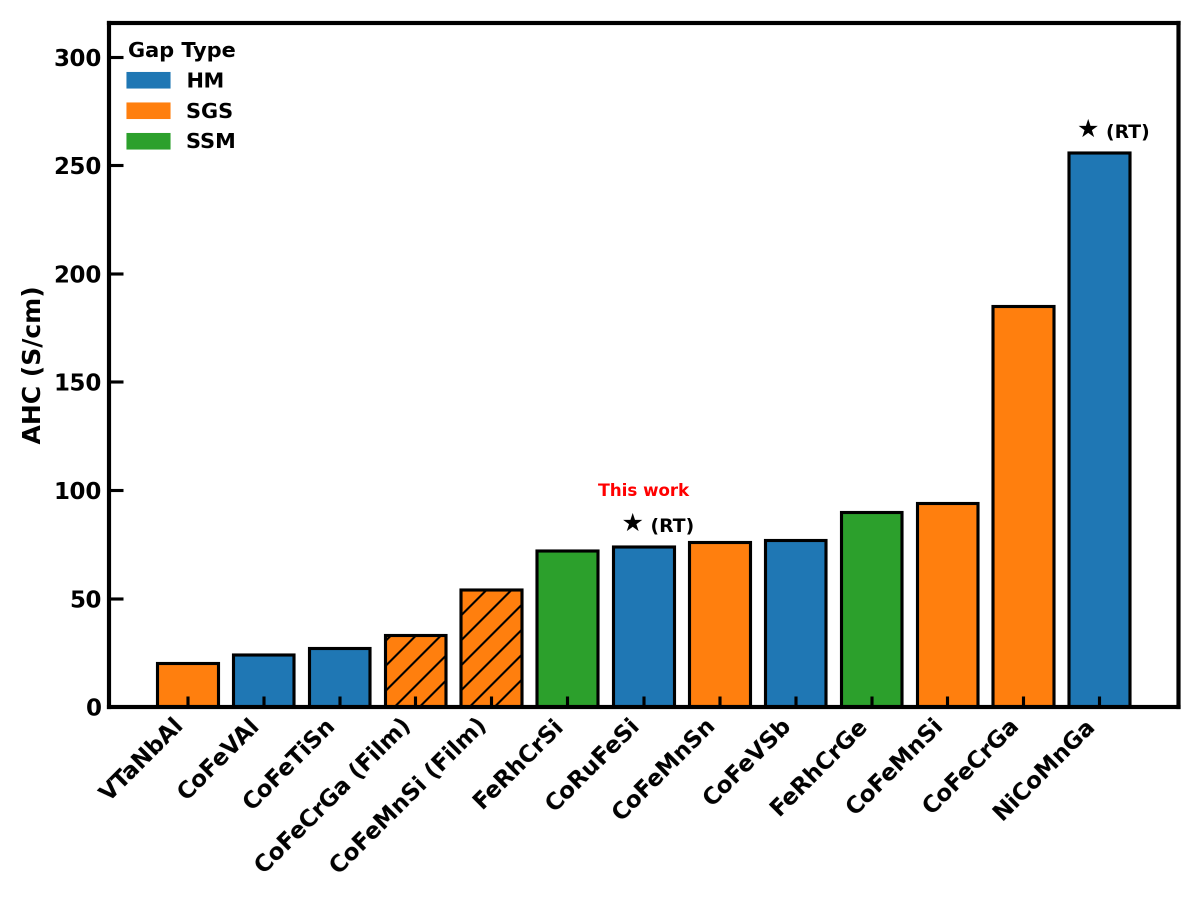}
    \caption{AHC values for various quaternary Heusler systems compiled from the literature (HM- Half-metal, SGS-Spin Gapless Semiconductor, SSM-Spin Semi Metal). Striped bars indicate thin-film samples, solid bars denote bulk systems, and the star marks the AHC measured at room temperature (RT) \cite{CoFeVSb_AHE, CoFeVAl_AHE, CoFeVAl_AHE, CoFeTiSn_AHE,CoFeTiSn_HM, NiCoMnGa_AHE,NiCoMnGa_HM, CoFeCrGa_AHE, CoFeCrGabulk_AHE, CoFeMnSn_AHE, CoFeMnSibulk_AHE, VTaNbAl_AHE, FeRhCrSi_AHE, FeRhCrSi_AHE, FeRhCrGe_AHE, CoFeMnSithinfilm_AHE}. }
    \label{fig:ahcplot}
\end{figure}
Normally, the AHE in the material may arise due to different kinds of mechanisms, such as skew scattering, side jump, and Karpus-Luttinger (KL) mechanism \cite{SMIT195839,SMIT1955877,Berger1970sidejump,Karplus1954}. The first two contributions are considered as the extrinsic contribution, whereas the last one is considered as an intrinsic contribution. Skew scattering is an extrinsic process that occurs when impurities scatter electrons asymmetrically. It results in a Hall resistivity that scales linearly with longitudinal resistivity ($\rho_{xy}^{AHE} \propto \rho_{xx}$), which is typically dominant in clean systems.  A lateral displacement of charge carriers during scattering causes the side-jump mechanism, which is also extrinsic and scales quadratically ($\rho_{xy}^{AHE} \propto \rho_{xx}^2$). The KL mechanism is intrinsic and is derived from the electronic band Berry curvature, which also produces a quadratic dependence on $\rho_{xx}$ and an anomalous transverse velocity. Hence, to know about the above two processes in our material, scaling analysis is performed, which is shown in \hyperref[fig5]{Fig. 5(c)}. The value of the exponent comes out to be 1.83, which means the skew scattering effect is very small in this material. The side jump contribution in the ferromagnetic material is normally expressed as $(e^2/ha)(\epsilon_{SO}/E_F)$n, where h, a, $\epsilon_{SO}$, and $E_F$ are known as Planck's constant, lattice parameter, spin-orbit interaction, and Fermi energy, respectively. Usually, in these materials, the value for $\epsilon_{SO}/E_F$ takes the value 10$^{-2}$ \cite{Shukla2022,Shahi2022,Rastogi2025}, which means that the contribution from the side jump towards the anomalous Hall response can be neglected. Hence, in CoRuFeSi, the AHE arises mainly from the intrinsic Berry phase-driven KL mechanism.

\begin{figure*}[htbp]
\centering
\includegraphics[width=0.98\linewidth]{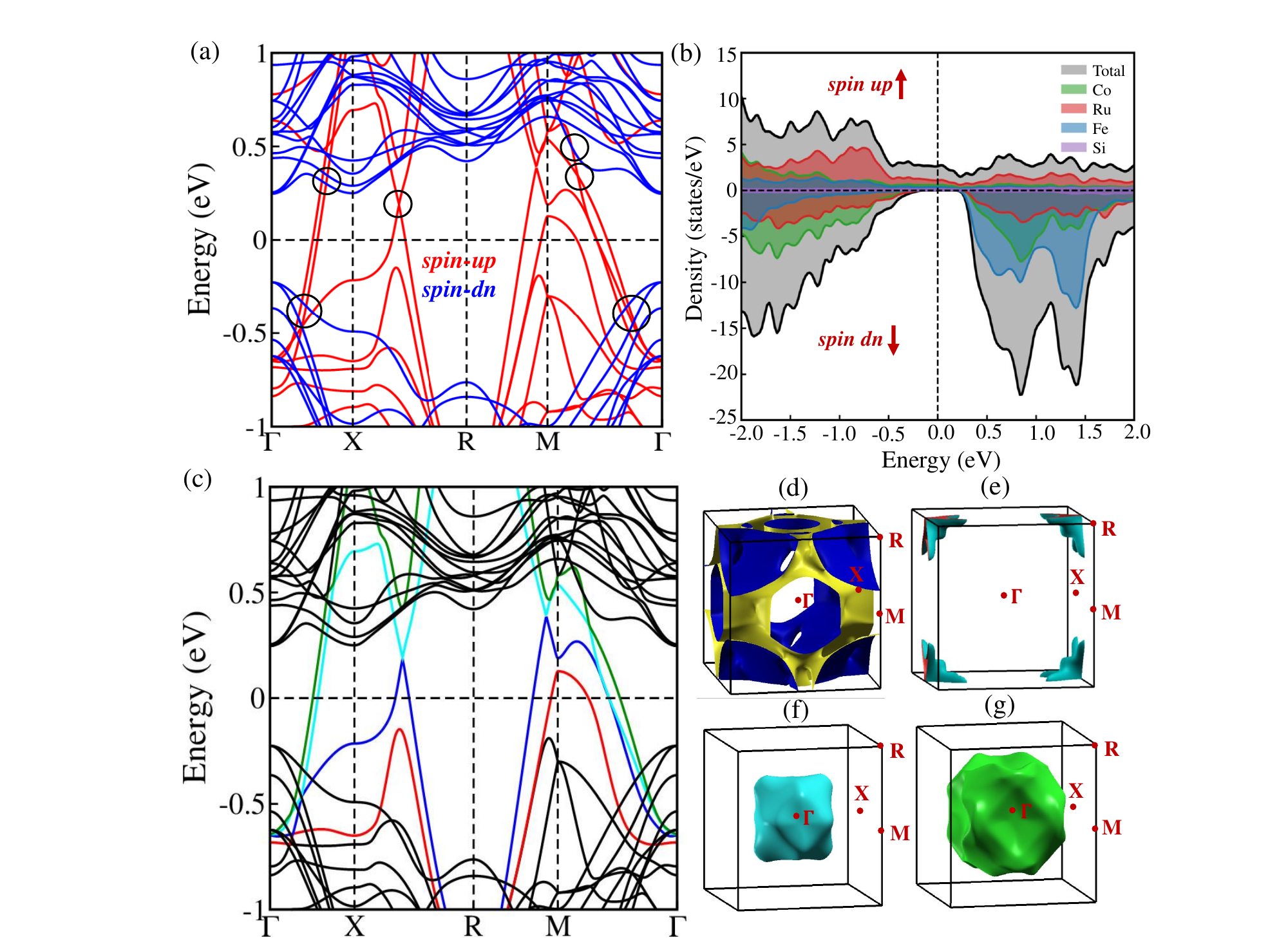}
\caption{(a) Spin-resolved electronic band structure of
CoRuFeSi without SOC. (b) Total and projected density of
states. (c) Electronic band structure of CoRuFeSi, with bands crossing the Fermi level highlighted in color. 
(d-g) Corresponding Fermi surface plots.}
\label{fs}
\end{figure*}
In order to understand the microscopic origin of the AHE in CoRuFeSi, it is crucial to examine how AHC depends on temperature and anomalous Hall resistivity.  The following conversion formula is used to perform this study:
\begin{equation}
    \sigma_{xy}=\frac{\rho_{xy}}{\rho_{xx}^2+\rho_{xy}^2}
\end{equation}

The Hall conductivity at various temperatures as a function of magnetic field is shown in \hyperref[fig5]{Fig. 5(d)}. Extrapolating the high-field Hall conductivity data to zero field along the y-axis allowed for the determination of the AHC ($\sigma_{xy}^A$).  \hyperref[fig5]{Fig. 5(e)} displays the temperature dependence of $\sigma_{xy}^A$ and $\sigma_{xx}^A$, demonstrating that AHC is almost temperature-invariant. This suggests the intrinsic origin of the AHE in CoRuFeSi \cite{nagaosa_review}.  The AHC is about 98 S/cm at 5 K and stays almost constant until room temperature, when it is about 95 S/cm.  The intrinsic contribution to AHE is particularly prominent when the longitudinal conductivity of the material ($\sigma_{xx}$) falls into the good metallic region, which is usually between 10$^4$ and 10$^6$ S/cm. The temperature-dependent $\sigma_{xx}$ is shown in the right scale of \hyperref[fig5]{Fig. 5(e)}, which have the value in the range $\sim$ 10$^4$ S/cm throughout the temperature region. These values are in the region of a good metallic regime, where the inherent Berry phase mechanism is the primary source of the AHE \cite{nagaosa_review}.
To extract the intrinsic contribution in the total AHC, we plotted the $\rho_{xy}^A$ vs $\rho_{xx}$ as shown in the \hyperref[fig5]{Fig. 5(f)} and fitted it with the equation :
\begin{equation}
    \rho_{xy}=\alpha\rho_{xx}+\sigma_{xy}^{int}\rho_{xx}^2
\end{equation}
where $\alpha$ and $\sigma_{xy}^{int}$ are known as the skew scattering coefficient and Berry phase driven intrinsic AHC, respectively. The values for the above two coefficients are found to be 0.003 and 74 S/cm, respectively. As both of the coefficient values are positive, one can say that both extrinsic and intrinsic mechanisms are in the same direction. The AHC value for different QHAs has been shown in the \hyperref[fig:ahcplot]{Fig. 6}. Although the AHC value in some QHAs is greater than that of CoRuFeSi, neither of them is at room temperature. We got the record high value of AHC in the case of a half-metallic ferromagnetic alloy at room temperature, which is mainly contributed by the intrinsic mechanism. Since the intrinsic AHE depends on the band structure of a material, to get a better understanding of the origin of intrinsic AHE, we carried out the first-principles calculation on CoRuFeSi.



\subsection{Theoretical results}
\subsubsection{Crystal and Electronic Structure Analysis}
Density functional theory (DFT) calculations were performed \textcolor{blue}{(see Supplementary Information)} to complement the experimental structural analysis. Three ordered crystal structures were considered, chosen to correspond exactly to the structural models employed in the FullProf Rietveld refinement discussed in the experimental section. Total-energy calculations identify the Type-YIII ordered structure as the ground-state configuration of CoRuFeSi, consistent with the experimental refinement. Furthermore, the optimized lattice parameter obtained for the Type-YIII structure is 5.79~\AA, in excellent agreement with the experimentally determined value of 5.78~\AA, as listed in \textcolor{blue}{Table~S1}.


\begin{figure}
\begin{center}
    \includegraphics[width=1.35\linewidth]{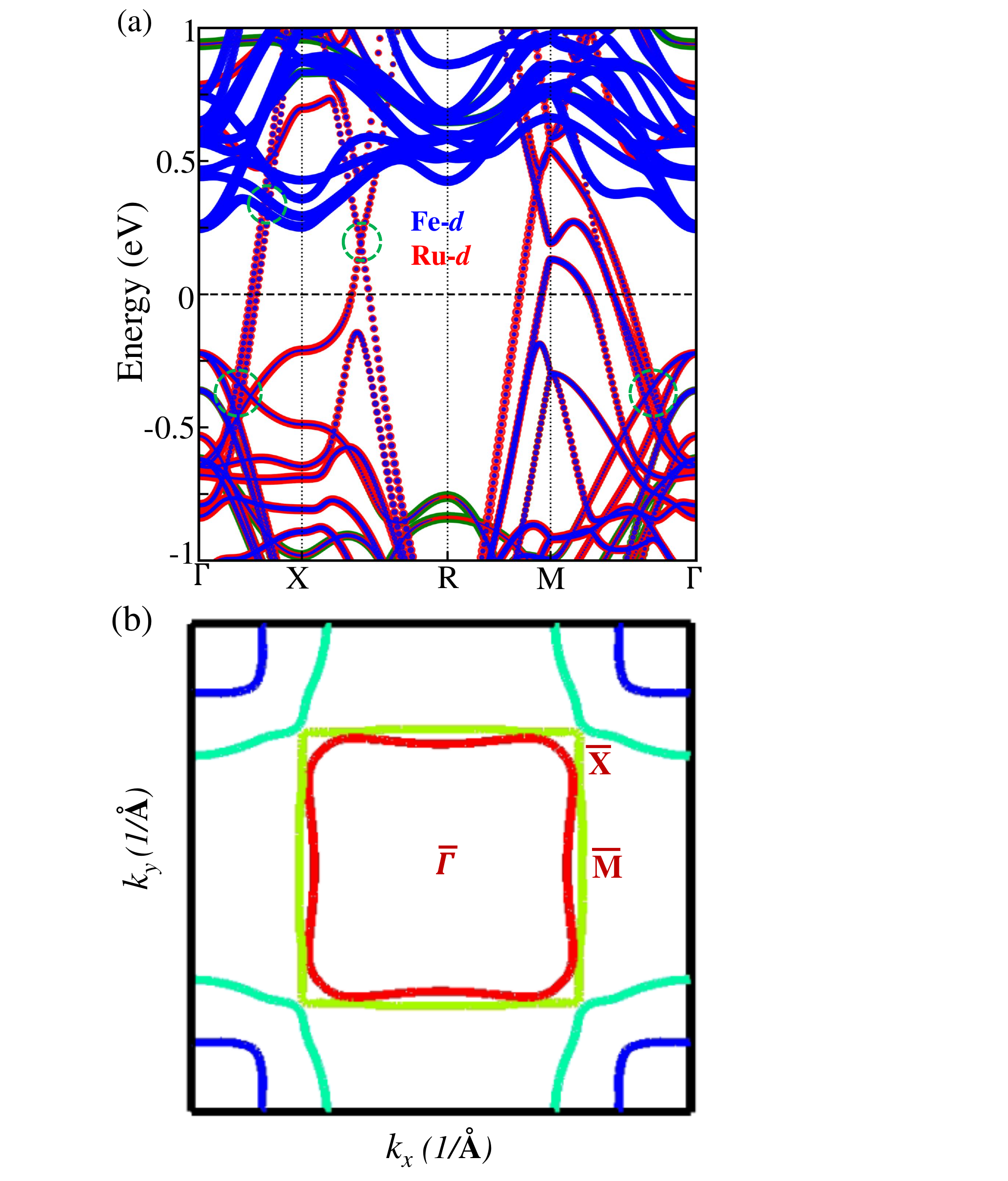}
    \caption{(a) Projected band structure showing orbital hybridization without SOC. (b) The nodal-ring dispersions along the (001) plane.}
    \label{nl}
\end{center}
\end{figure}

\begin{figure*}[htbp]
    \centering
    \includegraphics[width=1.15\linewidth]{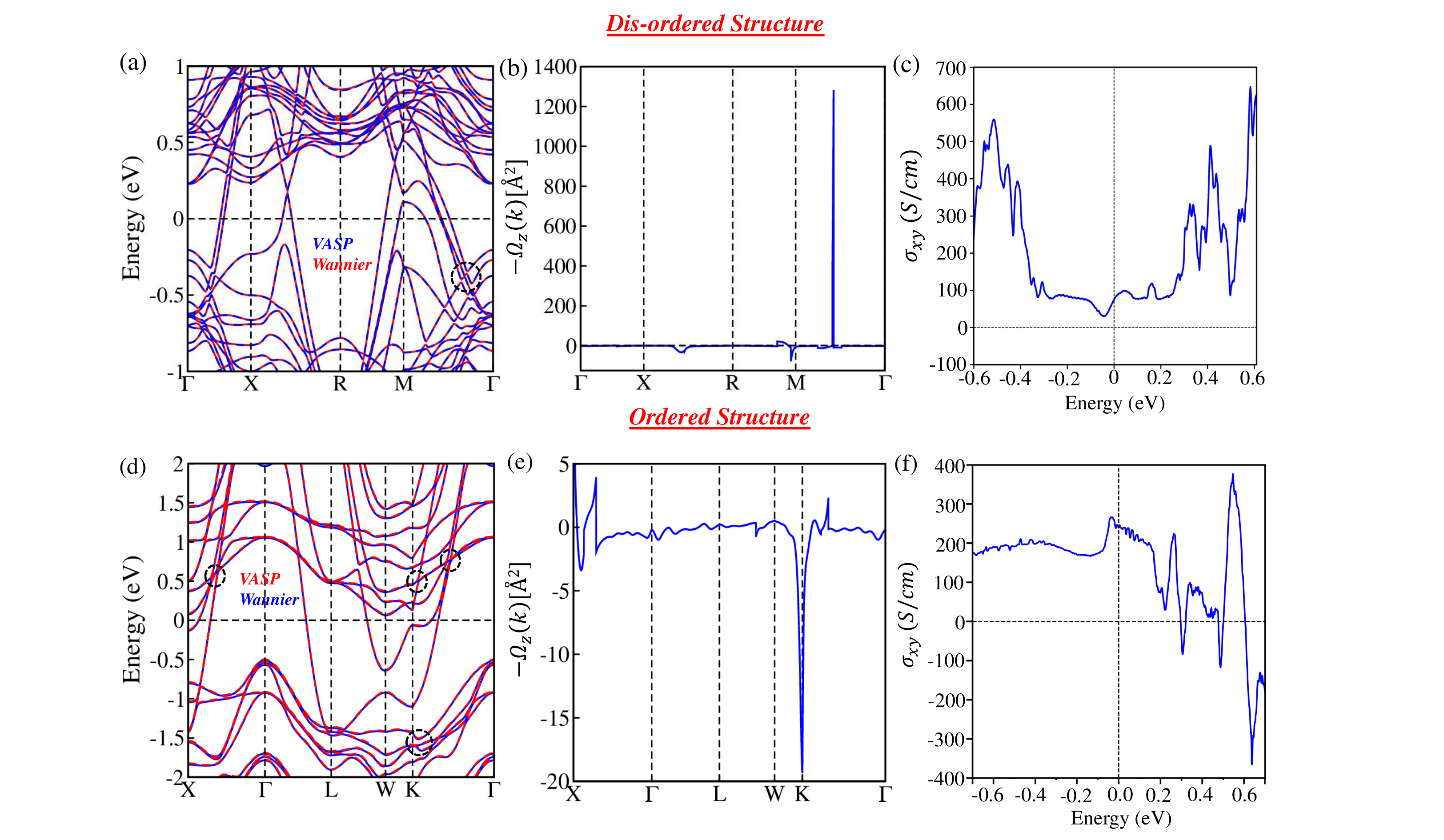}
    \caption{
    (a--c) Electronic structure and anomalous transverse properties of CoRuFeSi with SOC in the 50\% antisite-disordered configuration.
    (a) Comparison between Wannier90-interpolated bands (solid blue lines) and DFT bands from VASP (dashed red lines).  
    (b) Berry curvature plotted along the high-symmetry path.  
    (c) AHC for the disordered system.  
     (d--f) Corresponding results for the ordered structure:  
    (d) electronic band structure with SOC,  
    (e) Berry curvature, and  
    (f) Anomalous Hall Conductivity ($\sigma_{xy}$).}
    \label{ah2}
\end{figure*}

As disorder has been confirmed from the refinement of the PXRD data of \(\mathrm{CoRuFeSi}\), we performed first-principles calculations to gain further insight into the experimentally observed disordered structure. Antisite disorder between Co and Ru atoms was introduced at concentrations of 12.5\%, 25\%, and 50\%, as detailed in the \textcolor{blue}{Section 3 of the Supplementary Information}, using a \(2 \times 2 \times 2\) supercell. The corresponding density of states (DOS) indicates that the antisite-disordered system retains its half-metallic character. Furthermore, the calculated magnetic moment for disordered \(\mathrm{CoRuFeSi}\) is approximately \(5~\mu_B/\mathrm{f.u.}\), demonstrating that antisite disorder does not significantly affect the overall magnetization.

However, this large supercell contains 32 atoms, making Wannierization and the calculation of topological properties computationally expensive. Therefore, we proceeded with the conventional unit cell containing 16 atoms and introduced 50\% disorder by swapping two Co and Ru atoms. The spin-polarized electronic band structure of the CoRuFeSi  with 50\% disorder is illustrated in Fig.~\ref{fs}(a), where the spin-up and spin-down channels are shown in red and blue, respectively. The spin-up channel exhibits metallic behavior, whereas the spin-down channel shows semiconducting characteristics with a band gap of 0.47~eV, indicating the half-metallic nature of the system, which is further supported by the density of states (DOS) plot shown in Fig.~\ref{fs}(b). The atom-resolved projected density of states (PDOS) indicates that the electronic states forming both the valence and conduction bands are predominantly contributed by Co, Ru, and Fe atoms. In particular, Co states dominate the valence-band region, whereas Fe states mainly contribute to the conduction band, with a negligible contribution from Si. Notably, the density of states at the Fermi level is largely governed by Ru, Co, and Fe orbitals, which are expected to play a key role in determining the electronic transport and magnetic properties, as commonly observed in half-metallic systems \cite{Galanakis_HM}. To experimentally confirm the half-metallic character and the presence of disorder in the compound, we performed longitudinal resistivity measurements, as discussed in \textcolor{blue}{the Supplementary Section}.
  
The Fermi-surface plots for the four bands crossing the Fermi level ($E_F$) in CoRuFeSi without spin–orbit coupling (SOC), shown in Fig.~\ref{fs}(d–g), reveal four distinct pockets along the high-symmetry path $\Gamma$–$X$–$R$–$M$–$\Gamma$, corresponding to the four bands crossing $E_F$ shown in Fig.~\ref{fs}(c). Among these, the first pocket forms an open Fermi surface due to band crossings along the $X$–$R$ and $R$–$M$ directions, whereas the second pocket originates from a band crossing along the $R$–$M$–$\Gamma$ path. The last two pockets exhibit a closed, contour-like Fermi surface, corresponding to band crossings along the $\Gamma$--$X$ and $M$--$\Gamma$ directions. As these bands cross $E_F$ twice, transitioning from the valence band to the conduction band and vice versa, the system  hosts both electron- and hole-like charge carriers, with a predominance of hole-like pockets, in agreement with experimental observations. Similar multiple Fermi pockets have been reported in other Heusler compounds in the literature \cite{Halder2025CoMnVAlCoMnCrSi, Acharya2024Cr2FeCoAs2, Patel2022TbPtBi}.

\subsubsection{Topological Nodal-line Features}
As discussed above, the band-structure calculations reveal several notable features. In particular, we identify prominent band-crossing points along the $\Gamma$--$X$ and $M$--$\Gamma$ high-symmetry paths, marked by black circles in Fig.~\ref{fs}(a), which are indicative of potential topological nodal-line features in this system. The projected band structure, shown in Fig.~\ref{nl}(a), highlights the orbital hybridization between Fe-$d$ and Ru-$d$ states along the $\Gamma$--$X$ and $M$--$\Gamma$ directions. The dashed green circles denote the band crossing points characterized by opposite mirror eigenvalues of $+1$ and $-1$. Upon inclusion of spin--orbit coupling (SOC), this hybridization leads to a band inversion, thereby confirming the nontrivial topological nature of the nodal-line features. The presence of these nodal-line features is further corroborated by the two-dimensional Fermi-surface (2D FS) plots, which exhibit two closed, square-shaped nodal-ring dispersions, as shown in Fig.~\ref{nl}(b).
Upon including SOC with the magnetization along the [001] direction, the mirror symmetries that protect the nodal lines are broken, lifting the degeneracy at the crossing points and opening gaps at these crossings, as shown in Fig.~\ref{ah2}(a).

\subsubsection{Anomalous Hall Conductivity}
To verify the experimental findings and to understand the physical origin of the observed anomalous Hall response, we performed AHC calculations using the Wannier90 code. In this approach, Wannier functions were constructed from the DFT band structure, and an excellent agreement between the DFT and Wannier-interpolated bands was achieved, as shown in Fig.~\ref{ah2}(a). This accurate Wannier representation of the electronic structure provides a reliable foundation for evaluating Berry curvature and Anomalous Hall Conductivity (AHC).

The topological phase of a material is closely linked to its band-structure characteristics, particularly the Berry curvature (BC), which plays a central role in anomalous transport phenomena such as the AHC. Both the BC and the resulting AHC can be effectively tuned by modifying the electronic band structure and the underlying crystalline symmetries, independent of the net magnetization. In the presence of SOC, the magnetization direction—especially along the [001] axis—modifies the system’s symmetry, thereby inducing a finite Berry curvature $\Omega_z(\mathbf{k})$ near the Fermi energy $E_F$. Using the maximally localized Wannier functions (MLWFs), the BC is computed via the Kubo formula~\cite{Yao2004}:

\begin{equation}
\Omega_{n}^z(\mathbf{k}) = -2\, \mathrm{Im} \sum_{m \neq n}
\frac{\langle \psi_{n\mathbf{k}} | \hat{v}_{x} | \psi_{m\mathbf{k}} \rangle 
      \langle \psi_{m\mathbf{k}} | \hat{v}_{y} | \psi_{n\mathbf{k}} \rangle}
     {\left[E_m(\mathbf{k}) - E_n(\mathbf{k})\right]^2},
\label{bc}
\end{equation}

where $E_n(\mathbf{k})$ is the band energy corresponding to the $n$-th eigenstate $\psi_{n\mathbf{k}}$, $\mathbf{k}$ is the crystal momentum, $n$ represents the band index, and $\hat{v}_{x,y}$ denote the velocity operators. The intrinsic AHC, represented by the transport coefficient $\sigma_{xy}$, is then obtained by integrating the Berry curvature over all occupied states in the Brillouin zone:

\begin{equation}
\sigma_{xy}^z = -\frac{e^2}{\hbar}
\int_{BZ} \frac{d^3\mathbf{k}}{(2\pi)^3}
\sum_{n} f(\mathbf{k})\, \Omega_{n}^z(\mathbf{k}),
\end{equation}

where $f(\mathbf{k})$ is the Fermi–Dirac distribution function. The broken nodal lines with the inclusion of SOC, give rise to pronounced peaks along the $M$--$\Gamma$ direction and shallow valleys along the $X$--$R$ and $R$--$M$ directions in the Berry curvature, as shown in Fig.~\ref{ah2}(b). These SOC-induced gapped crossing points are marked by black dotted circles in Fig.~\ref{ah2}(a). 
Interestingly, the theoretically calculated anomalous Hall conductivity (AHC) for the disordered CoRuFeSi system is
$74$~S/cm, which is in excellent agreement with the experimental value of
$74$~S/cm, as shown in Fig.~\ref{ah2}(c).


To investigate the topological and anomalous transverse properties of the system in its ordered phase, as well as its experimental feasibility, we performed a detailed analysis of the ordered structure, revealing the intriguing features that arise under perfect atomic ordering. The interesting non-trivial topological characteristics of the ordered system are discussed in the \textcolor{blue}{Fig.~S6}, and the calculated AHC for the ordered phase is discussed here. The calculated AHC for ordered CoRuFeSi at the Fermi level is $238$~S/cm, as shown in Fig.~\ref{ah2}(f), indicating a robust intrinsic transport response even in the absence of external perturbations such as doping or Fermi level shifts. However, moving away from the Fermi level, the AHC value decreases, reflecting its sensitivity to the position of the chemical potential.
Moreover, when comparing the ordered and disordered phases, although the disordered system exhibits sharp peaks in the Berry curvature, the overall AHC is reduced because these peaks originate from SOC-induced gaps at avoided crossing points located away from the Fermi level (E$_F$). In contrast, for the ordered structure, peaks in the anomalous Hall conductivity (AHC) arise from SOC-induced gaps near the Fermi level, leading to a higher AHC at $E_F$. Consistent with this behavior, the narrow and sharp peak of the Berry curvature in disordered CoRuFeSi, shown in Fig.~\ref{ah2}(b), reflects a reduced contribution across $k$-space, resulting in a lower anomalous Hall conductivity (AHC) compared to the ordered phase. These observations confirm that the experimentally observed suppression of AHC in the disordered system arises from antisite disorder–induced modifications of the Berry curvature, while the half-metallic character of the system remains largely preserved.
 Our theoretical analysis further demonstrates that disorder modifies the Berry curvature distribution and consequently suppresses the intrinsic AHC, in agreement with recent reports indicating that disorder generally reduces the AHC relative to ordered systems~\cite{sakuraba2020, mende2021}.\\
Taken together with our experimental findings, these results reveal that the impact of disorder on Berry curvature is not universal but depends sensitively on disorder-induced changes to the electronic structure, which vary among different Heusler compounds due to their distinct valence electron configurations. When disorder shifts band crossings or avoided crossings away from the Fermi level, the AHC is strongly diminished. Overall, our study establishes a framework for the systematic investigation of the AHE in disordered Heusler compounds, particularly CoRuFeSi. Similar methodologies for calculating the AHC in disordered systems have been adopted in previous works~\cite{Shukla2022, Shahi2022, Rastogi2025}. These findings underscore that modifications to the electronic structure are the primary factor governing the observed AHC in disordered CoRuFeSi.


\section{CONCLUSIONS}
We have synthesized the quaternary Heusler alloy CoRuFeSi by the arc-melting technique, and x-ray diffraction confirms the LiMgPdSn-type structure with approximately 50\% Co--Ru antisite disorder. CoRuFeSi exhibits soft ferromagnetic behavior at room temperature. Hall resistivity measurements reveal a clear anomalous Hall contribution, establishing the presence of the anomalous Hall effect. The anomalous Hall conductivity is nearly temperature independent, with a total value of $98~\mathrm{S/cm}$, of which the intrinsic contribution is $74~\mathrm{S/cm}$.
First-principles calculations identify CoRuFeSi as a topologically nontrivial nodal-line semimetal. For the ordered phase, the intrinsic anomalous Hall conductivity is calculated to be $238~\mathrm{S/cm}$, which is significantly higher than the experimentally observed value. Incorporation of 50\% Co--Ru antisite disorder suppresses the Berry curvature and reduces the anomalous Hall conductivity, bringing the calculated value into close agreement with experiment while preserving the half-metallic character. These results highlight CoRuFeSi as a disorder-tolerant half-metallic ferromagnet with a sizable intrinsic anomalous Hall effect at room temperature, underscoring its potential for spintronic and Hall-based device applications.


\section*{ACKNOWLEDGMENTS}
M.P. would like to acknowledge NIT Andhra Pradesh for the fellowship. T.P. would like to acknowledge UGC DAE CSR (Grant No. CRS/2021-22/02/487) and SERB (Grant No. CRG/2022/008197) for their financial support. SM acknowledges DST, India. S.S.P. and V.K. sincerely acknowledge the National Supercomputing Mission (NSM) for providing computational resources on ‘PARAM SEVA’ at IIT Hyderabad. S.S.P. acknowledges  DST-INSPIRE for a research fellowship and V.K. expresses gratitude for the support provided through the DRDO Project No. ERIP/ER/202312003/M/01/1853.

\bibliography{CoRuFeSi}

\end{document}


\begin{titlepage}
\centering
\renewcommand{\thefootnote}{\fnsymbol{footnote}}

\vspace*{4cm}

{\Huge\bfseries Supplementary Information\par}
\vspace{1.5cm}

{\LARGE\bfseries 
Room temperature intrinsic anomalous Hall effect in disordered half-metallic ferromagnetic quaternary Heusler alloy CoRuFeSi
\par}
\vspace{2.5cm}
{\normalsize
\textbf{Manikantha Panda},$^{1}$
\textbf{Sonali S. Pradhan},$^{2}$
\textbf{Prabuddha Kant Mishra},$^{3}$\\
\textbf{Alapan Bera},$^{4}$
\textbf{Rosni Roy},$^{5}$
\textbf{Rajib Mondal},$^{5}$
\textbf{Soumik Mukhopadhyay},$^{4}$\\
\textbf{V.~Kanchana},$^{2}$\footnote{E-mail: kanchana@phy.iith.ac.in}
\textbf{Tapas Paramanik}$^{1}$\footnote{E-mail: tapas.phys@nitandhra.ac.in}
\par}
\vspace{1cm}

{\small
$^{1}$Department of Physics, National Institute of Technology Andhra Pradesh,\\
Tadepalligudem 534101, India\\[0.3cm]

$^{2}$Department of Physics, Indian Institute of Technology Hyderabad,\\
Kandi 502285, Sangareddy, Telangana, India\\[0.3cm]

$^{3}$Current affiliation: Institute of Low Temperature and Structure Research,\\
Polish Academy of Sciences, Ok\'olna 2, 50--422 Wroc\l aw, Poland\\[0.3cm]

$^{4}$Department of Physics, Indian Institute of Technology Kanpur, Kanpur 208016, India \\[0.3cm]

$^{5}$UGC-DAE Consortium for Scientific Research, Kolkata Centre,\\
Sector III, LB-8, Salt Lake, 700106, West Bengal, India
\par}

\vfill

\end{titlepage}

\setcounter{page}{2}
\pagestyle{fancy}
\fancyhf{}
\rhead{\thepage}


\section{Crystal Structure}
\begin{figure*}
    \centering
    \includegraphics[width=1\linewidth]{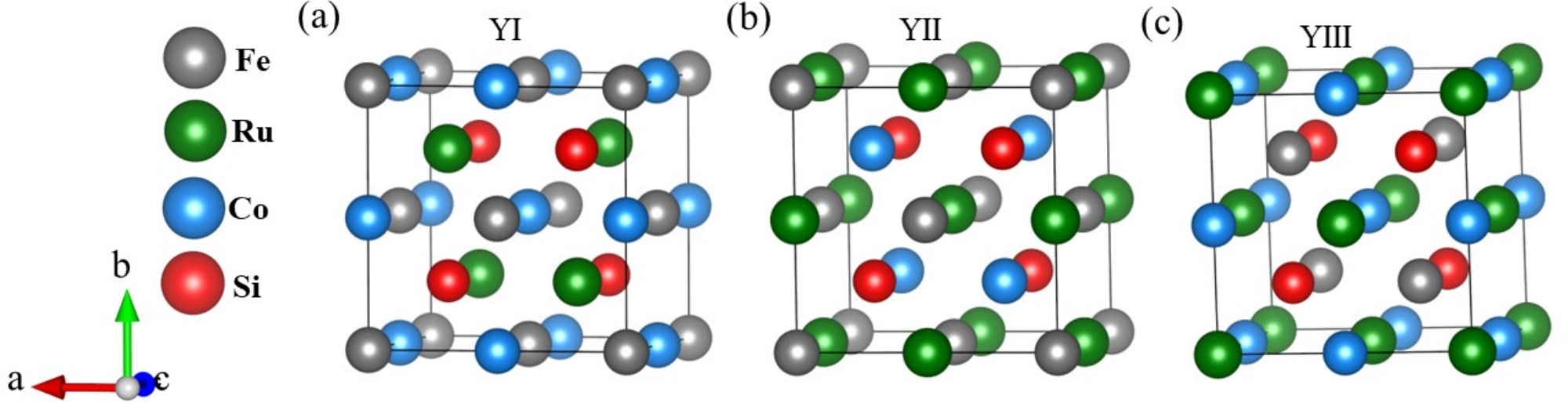}
    \caption{Crystal structure representations of the YI, YII, and YIII ordered phases of CoRuFeSi.}
    \label{cry}
\end{figure*}

CoRuFeSi belongs to the class of quaternary Heusler compounds of the type XX$'$YZ and crystallizes in the LiMgPdSn-type structure with space group $F\bar{4}3m$ \cite{Bainsla2016}. In 
 this equiatomic quaternary Heusler alloys (EQHAs), there are four distinct Wyckoff positions: $4a$~$(0, 0, 0)$, $4b$~$(0.5, 0.5, 0.5)$, $4c$~$(0.25, 0.25, 0.25)$, and $4d$~$(0.75, 0.75, 0.75)$~\cite{Graf2011, Neibecker2017}. Typically, the main group element $Z$ occupies the $4d$ site, while the transition metals $X$, $X'$, and $Y$ occupy the $4a$, $4b$, and $4c$ sites, respectively. Three distinct ordered configurations (YI, YII, and YIII) can be realized, depending on whether the $4c$ Wyckoff position $(0.25,\,0.25,\,0.25)$ is occupied by $X'$, $Y$, or $X$, respectively. For CoRuFeSi, theoretical calculations reveal that the YIII configuration has the lowest energy and is therefore the most energetically favorable, as shown in \textcolor{blue}{Table~S\ref{ta1}}. In the YIII configuration, Co, Ru, Fe, and Si are located at the $4c$~$(0.25,\,0.25,\,0.25)$, $4a$~$(0,\,0,\,0)$, $4b$~$(0.5,\,0.5,\,0.5)$, and $4d$~$(0.75,\,0.75,\,0.75)$ Wyckoff sites, respectively, as illustrated in \textcolor{blue}{Fig.~S\ref{cry}}. The optimized lattice parameters are $a = b = c = 5.79$~\AA{}, which are in excellent agreement with the experimental values of $a = b = c = 5.78$~\AA{} \cite{Benkaddour2016CoRuFeZ,Seema2019CoRuFeSi}.

 

\begin{table}[h!]
\centering
\caption{Ordered structure types, their atomic positions, and the corresponding energies.}
\label{tab:structure_energy}
\begin{tabular}{|c|p{8cm}|c|}
\hline
\textbf{Structure Type} & \textbf{Atomic Positions} & \textbf{Energy (eV)} \\
\hline
YI   & X\, (0,0,0), X$'$\,(0.25, 0.25, 0.25), Y\, (0.5, 0.5, 0.5), Z\, (0.75, 0.75, 0.75) & $-121.77$ \\
\hline
YII  & X\, (0,0,0), X$'$\,(0.5, 0.5, 0.5), Y\, (0.25, 0.25, 0.25), Z\, (0.75, 0.75, 0.75) & $-123.85$ \\
\hline
YIII & X\, (0.25, 0.25, 0.25), X$'$\,(0,0,0), Y\, (0.5, 0.5, 0.5), Z\, (0.75, 0.75, 0.75) & $-125.99$ \\
\hline
\end{tabular}
\label{ta1}
\end{table}

\begin{table}[htb]
\centering
\normalsize 
\caption{Relaxed lattice parameter ($a_0$), atom-projected magnetic moments ($\mu_\mathrm{B}$) for Co, Ru, and Fe atoms, total magnetic moment per formula unit ($\mu_\mathrm{B}$).}
\label{tab:crms_configs}
\begin{tabular}{lcccccc}
\hline
Type & $a_0$ (\AA) & $m_\mathrm{Co}$ & $m_\mathrm{Ru}$ & $m_\mathrm{Fe}$ & Total  \\
     &             & ($\mu_\mathrm{B}$) & ($\mu_\mathrm{B}$) & ($\mu_\mathrm{B}$) & ($\mu_\mathrm{B}$)  \\
\hline
YIII   & 5.79 & 1.50 & 0.42  & 3.14  & 5.06  \\

\hline
\end{tabular}
\end{table}

\section{Magnetocrystalline anisotropy}
To compare with the experiment findings, first-principles calculations were performed. The calculated magnetic moment associated with the Fe atoms is $5.0~\mu_{\mathrm{B}}$. To determine the magnetic easy axis, total-energy calculations were carried out for several spin orientations, specifically along the [001], [010], [100], [110], [011], [101], and [111] directions. The [001] direction is found to have the lowest energy as shown in \textcolor{blue}{Table~S3}, identifying it as the easy axis and indicating that the preferred spin orientation lies along the $c$ axis.

\begingroup
\color{blue}
\begin{table}
	\centering
	\caption{The relative energies of various spin orientations in the FM state, taking the lowest-energy configuration (001) as the reference.}
	\setlength{\tabcolsep}{3pt}
	\renewcommand{\arraystretch}{1.2}
	\begin{tabular}{@{}l *{7}{c}@{}}
		\hline\hline
		Configuration & 001 & 100 & 010 & 110 & 011& 101& 111 \\  
		\hline
		Energy/f.u. ($\mu$eV) & 
		0 & 7.98 & 11.77 & 10.24 & 2.27 & 4.86 & 26.02 \\  
		\hline
	\end{tabular}
	\label{taS3}
\end{table}
\endgroup

\section{Robust half-metallicity in disordered lattice}

\begin{figure*}
    \centering
    \includegraphics[width=0.8\linewidth]{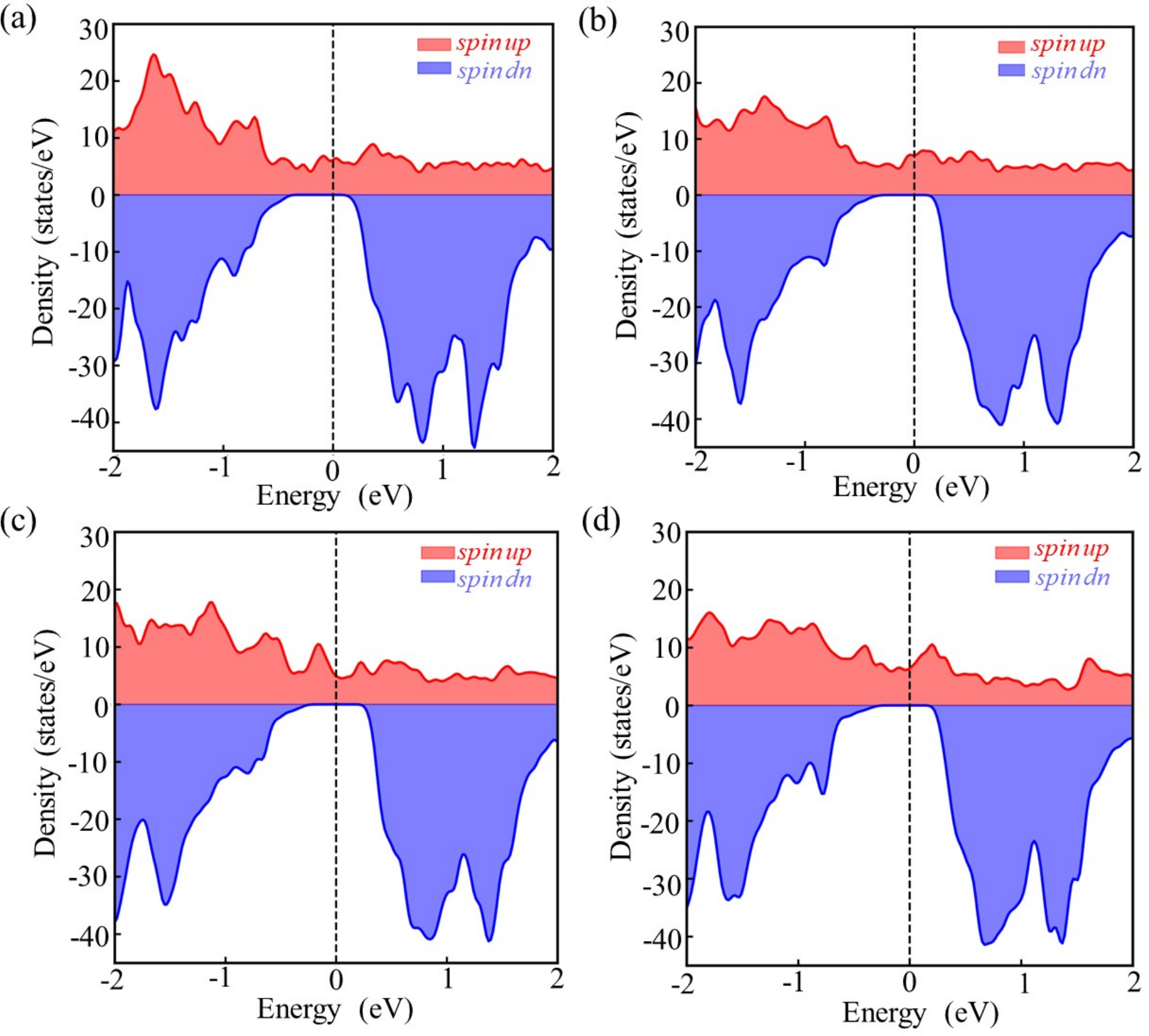}
    \caption{Spin-resolved density of states (DOS) for CoRuFeSi with (a) 12.5\%, (b) 25\%, (c) 37.5\%, and (d) 50\% Co--Ru swap disorder.}
    \label{dis}
\end{figure*}
To gain insight into the electronic and magnetic properties of experimental structure, we carried out first-principles calculations on disordered configurations by introducing a random distribution of Co and Ru atoms at the $4c$ and $4a$ Wyckoff positions. For this purpose, a $2 \times 2 \times 2$ supercell containing 32 atoms (corresponding to 8 formula units) was constructed. The symmetrically equivalent positions allow us to swap any of the eight possible positions of Co with any of the eight choices of Ru.
Within this framework, the exchange of a single Co–Ru pair corresponds to a 12.5\% swap disorder, while exchanging two, three, and four Co–Ru pairs corresponds to 25\%, 37.5\%, and 50\% disorder levels, respectively. A 50\% swap disorder thus represents a disordered structure analogous to the $L2_{1}$-type phase. The spin-polarized DOS corresponding to varying disorder concentrations is presented in \textcolor{blue}{Fig. S\ref{dis}(a-d)}. These results clearly indicate that the half-metallic character of CoRuFeSi remains robust against Co–Ru swapping disorder. Furthermore, the total magnetic moment remains nearly unchanged compared to that of the ideal, ordered structure.

To further understand the magnetic behavior, we evaluated the local magnetic moments at and around the atomic sites, with the results summarized in \textcolor{blue}{Table~S\ref{ta2}}. It is observed that Co atoms occupying Ru sites exhibit a slight enhancement in their magnetic moments. Importantly, despite the presence of disorder, the local chemical environment surrounding the magnetic atoms remains largely unaffected, with each atom still coordinated by eight Co atoms as in the pristine structure. Only minor bond-length variations were observed, primarily due to local relaxation effects. Consequently, the magnetic exchange interactions are minimally influenced, resulting in only slight modifications to the local and total magnetic moments. Overall, the Co–Ru disordered configurations retain magnetic characteristics that closely resemble those of the ideal CoRuFeSi structure.

\begin{table}[htbp]
\centering
\caption{Magnetic moments ($\mu_B$) of the perfect and disordered CoRuFeSi supercell ($2 \times 2 \times 2$). Co1--Co8 denote the eight Co atoms in the supercell, while $X_d$ indicates defect atoms. Co$_\mathrm{Ru}$ represents the magnetic moment of a Co atom at Ru site, and Ru$_\mathrm{Co}$ represents that of a Ru atom at Co site.}
\resizebox{\textwidth}{!}{%
\begin{tabular}{|c|c|c|c|c|c|c|}
\hline
System & $Co_1-Co_8$ atoms & $Ru_1-Ru_8$ atoms & $Fe_1-Fe_8$ atoms & $X_d$ & $m_\text{total}$ ($\mu_B$/cell) & $m_\text{total}$ ($\mu_B$/f.u.) \\
\hline
Perfect (no swap) & 1.512 & 0.419 & 3.155 & -- & 40.619 & 5.00 \\
\hline
\begin{tabular}[c]{@{}c@{}}Co-Ru\\ swap\\ (12.5\%)\end{tabular} & \begin{tabular}[c]{@{}c@{}}0.97, 0.96,\\ 0.98, 0.97,\\ 0.96, 1.00,\\ 0.98\end{tabular} & \begin{tabular}[c]{@{}c@{}}0.08, 0.05,\\ 0.08, 0.18,\\ 0.09, 0.05,\\ 0.02\end{tabular} & \begin{tabular}[c]{@{}c@{}}2.99, 2.92,\\ 3.02, 2.90,\\ 2.92, 3.00,\\ 2.92, 3.00\end{tabular} & \begin{tabular}[c]{@{}c@{}}Co$_\text{Ru}$: 1.08,\\ Ru$_\text{Co}$: $-$0.16\end{tabular} & 40.578 & 5.00 \\
\hline
\begin{tabular}[c]{@{}c@{}}Co-Ru\\ swap\\ (25.0\%)\end{tabular} & \begin{tabular}[c]{@{}c@{}}1.534, 1.534,\\1.526, 1.526,\\  1.457, 1.457
\end{tabular} & \begin{tabular}[c]{@{}c@{}}   0.351,
     0.351,\\     0.372,
     0.372,\\
     0.460,
     0.460\end{tabular} & \begin{tabular}[c]{@{}c@{}}  3.184,
   3.184,\\
      3.182,
   3.182,\\
   3.187,
   3.187,\\
  3.146,
   3.146,\\
 3.178, 3.100\end{tabular} & \begin{tabular}[c]{@{}c@{}}Co$_\text{Ru}$:1.524, 1.524

\\ Ru$_\text{Co}$:    0.357   0.357\end{tabular} & 40.528 & 5.00 \\
\hline
\begin{tabular}[c]{@{}c@{}}Co-Ru\\ swap\\ (37.0\%)\end{tabular} & \begin{tabular}[c]{@{}c@{}} 1.535,\\ 1.512, 1.512,\\ 1.481, 1.488\end{tabular} & \begin{tabular}[c]{@{}c@{}}
0.320,\\ 0.354,  0.405,\\ 0.451,  0.432\end{tabular} & \begin{tabular}[c]{@{}c@{}}3.161,  3.186
,\\ 3.185, 3.185,\\  3.179,  3.179,\\ 3.178, 3.100\end{tabular} & \begin{tabular}[c]{@{}c@{}}Co$_\text{Ru}$: 1.523
, 1.527, 1.512\\ Ru$_\text{Co}$: 0.390
, 0.382, 0.410\end{tabular}  & 40.549 & 5.00 \\
\hline
\begin{tabular}[c]{@{}c@{}}Co-Ru\\ swap\\ (50\%)\end{tabular} & \begin{tabular}[c]{@{}c@{}} 1.507,  1.507,\\  1.507,  1.507\end{tabular} & \begin{tabular}[c]{@{}c@{}}0.390, 0.390,\\ 0.390, 0.390 \end{tabular} & \begin{tabular}[c]{@{}c@{}}3.174, 3.174,\\ 3.174, 3.174,\\ 3.174, 3.174,\\ 3.174, 3.174\end{tabular} & \begin{tabular}[c]{@{}c@{}}Co$_\text{Ru}$: 1.507,1.507, 1.507
, 1.507\\ Ru$_\text{Co}$: 0.390
, 0.390
, 0.390
, 0.390
\end{tabular} & 40.532 & 5.00 \\
\hline
\end{tabular}%
}
\label{ta2}
\end{table}
\begin{figure*}
    \centering
    \includegraphics[width=1\linewidth]{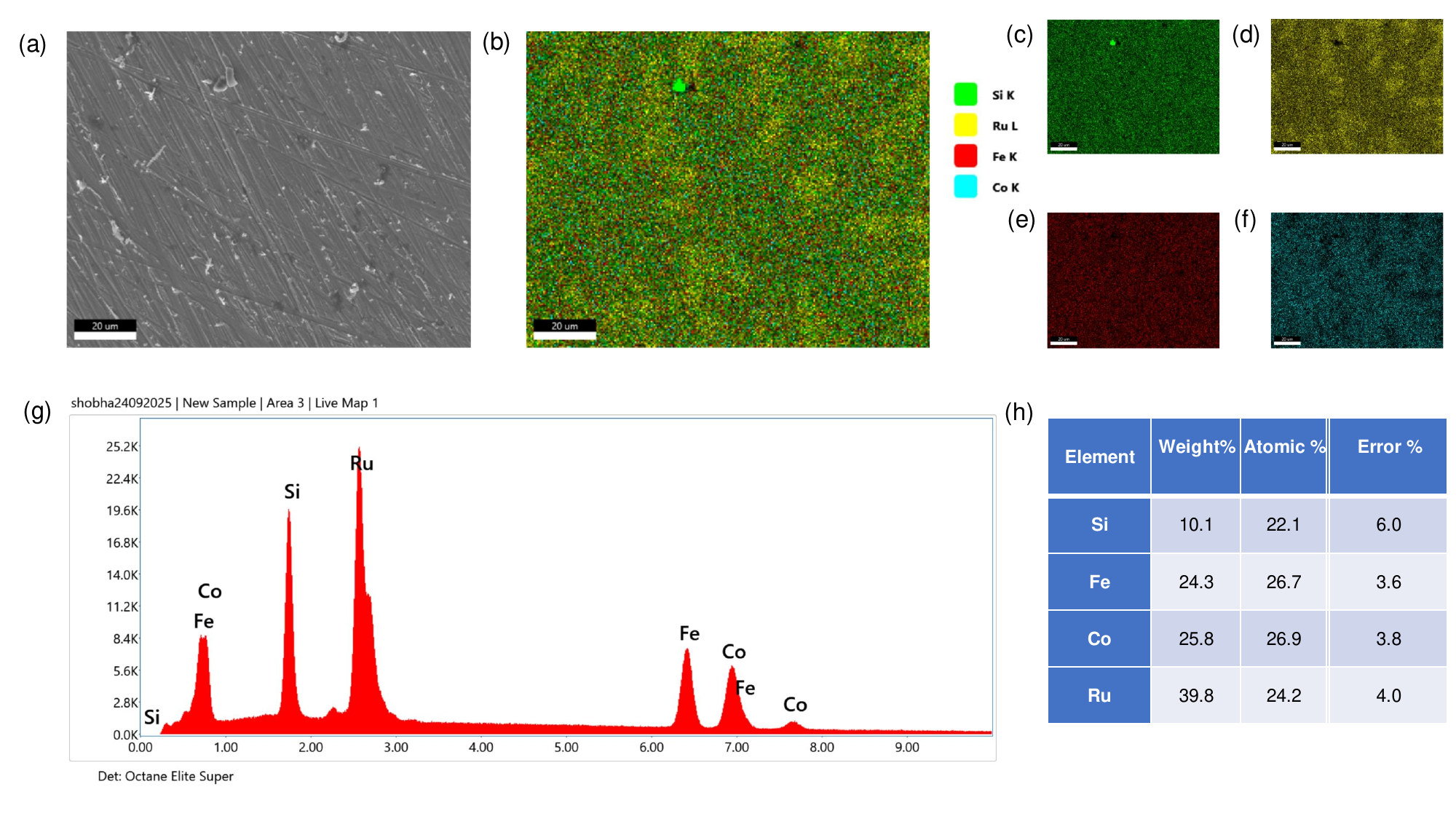}
    \caption{Energy dispersive analysis data for CoRuFeSi. (a) Scanning electron microscopy image. (b-f) mapping of the compound shows the homogeneity of it. (g-h) shows the elemental composition.}
    \label{edx1}
\end{figure*}


\section{Resistivity data }
Based on the slope of $\rho_{xx}(T)$ data, the features can be discussed separately in two distinct temperature regimes, i.e, (i) high temperature regime 75~K $< T<$ 300~K and (ii) low temperature regime 2~K$< T<$75~K.


The quasi-linear temperature dependence of high temperature $\rho_{xx}(T)$ can be fitted with following relation,

\begin{equation}\label{resist}
\rho{_{xx}} = \rho_0 + \rho_{e-p} T +\rho_{e-m}T^{2}
\end{equation}
where $\rho_0$ is the residual resistivity, $\rho_{e-p}$  is the electron-phonon scattering coefficient and $\rho_{e-m}$ is the electron-magnon scattering coefficient. The obtained values of the different scattering coefficients are given in the table \ref{tab1}.   It can be seen that the value for $\rho_{e-m}$ is two orders of magnitude less than that of the value of $\rho_{e-p}$, suggesting the dominance of electron-phonon scattering in the high temperature region. The less feasible value of $\rho_{e-m}$, which may be neglected, is the characteristic of half-metallic ferromagnetic alloys \cite{VENKATESWARA2020166536}. This is aligned with our theoretical calculation as discussed.

\begin{figure}
\begin{center}
    \includegraphics[width=1\linewidth]{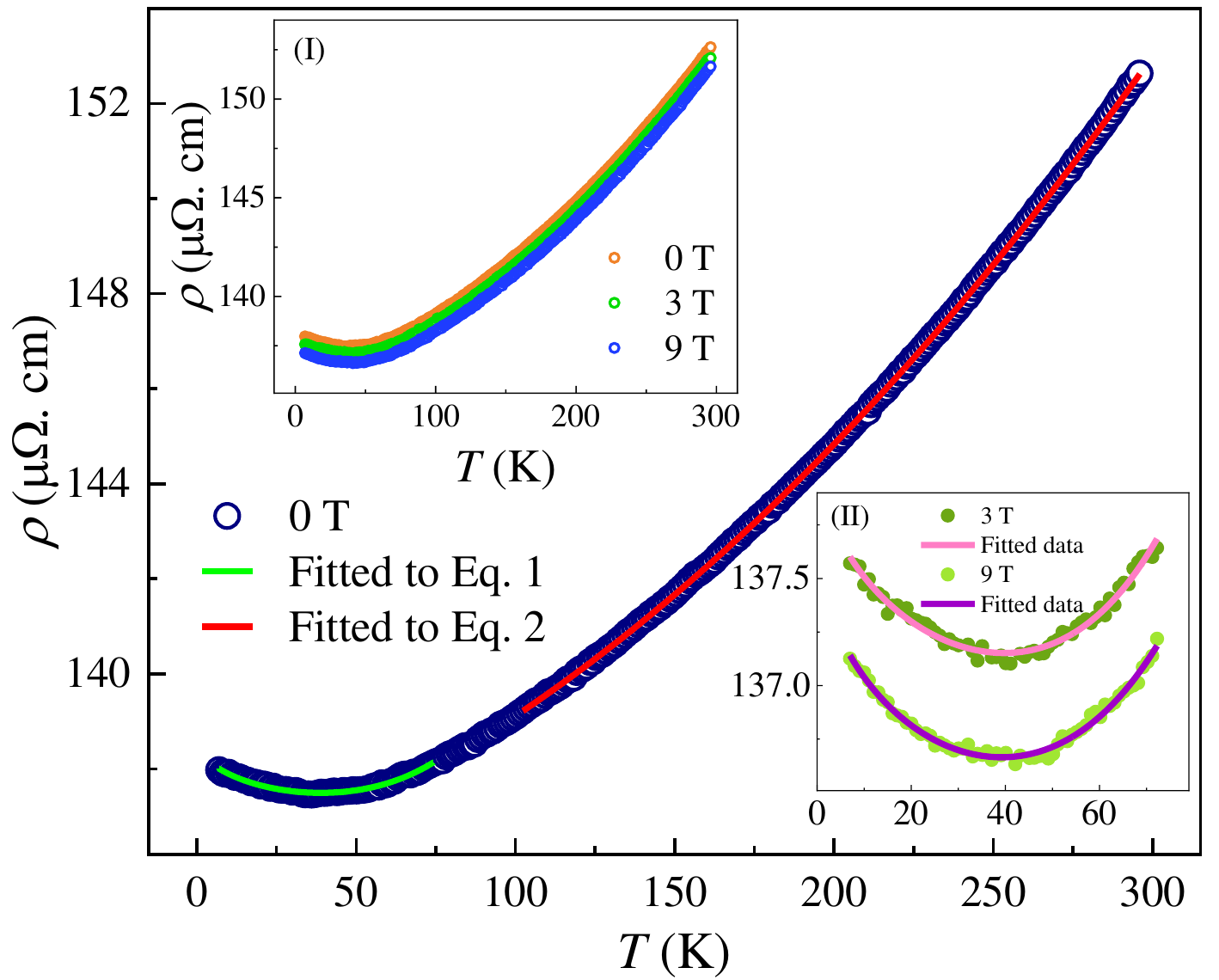}
    \caption{Temperature-dependent longitudinal resistivity fitted using Eq. 4 in the low-temperature range (T $<$ 75 K) and Eq. 5 in the higher temperature range (75 K $<$ T $<$ 300 K). Inset I displays the variation of resistivity under magnetic fields of 0 T, 3 T, and 9 T. Inset II shows the fitting of Eq. 4 to the resistivity data measured under 3 T and 9 T magnetic fields.}
    \label{fig3}
\end{center}
\end{figure}

The upturn in the low temperature $\rho_{xx}(T)$ data was measured at different magnetic fields, and the upturn feature remains invariable as shown in the inset of \hyperref[fig3]{Fig.3}.
It indicates it could be regarded as signature of EEI. To confirm this, low temperature regime is fitted with the following equation,

 \begin{equation}\label{resist}
\rho{_{xx}} = \rho_0 + \rho_{e-e}T^2 + \rho_{e-p} T^5 -\rho_{EEI}T^{1/2}
\end{equation}

where $\rho_0$ is the residual resistivity, $\rho_{e-e}$ is the electron-electron scattering coefficient (Fermi liquid behaviour), $\rho_{e-p}$  is the electron-phonon scattering coefficient of Bloch-Gruneisen (BG) form (at low temperature BG term takes the form $KT^5$), $\rho_{EEI}$ is the EEI scattering coefficient. Again, the same equation is fitted to the low temperature data of $\rho_{xx}(T)$ in the presence of magnetic fields of 3 T and 9 T, as shown in the inset II of \hyperref[fig3]{Fig. 3}. The well-fitting of the data suggests that in CoRuFeSi, the EEI is dominant in the low temperature region, which gives rise to the upturn in $\rho_{xx}(T)$.  
Similar low-temperature feature in resistivity is attributed to the presence of disorder in several Heusler alloys \cite{zhu2017anomalous,Garmroudi_2023,Ti2FeAl_koushik}, which was not prominent in the earlier report of CoRuFeSi \cite{BAINSLA2015631}.

\begin{table}[htbp]
\caption{Fitting parameters of the longitudinal resistivity in the low- and high-temperature regions.}
\label{tab1}
\centering
\scriptsize
\resizebox{\columnwidth}{!}{%
\begin{tabular}{lcccc}
\hline\hline
\multicolumn{5}{c}{\textbf{Low-$T$ region: $2 < T < 75$ K}} \\
\hline
Parameter & $\rho_{0}$ & $a$ & $b$ & $c$ \\
Units 
& ($\mu\Omega$\,cm) 
& ($10^{-4}$ $\mu\Omega$\,cm\,K$^{-2}$) 
& ($10^{-10}$ $\mu\Omega$\,cm\,K$^{-5}$) 
& ($\mu\Omega$\,cm\,K$^{-1/2}$) \\
Value & 138.61 & 2.19 & 1.47 & 0.23 \\
\hline
\multicolumn{5}{c}{\textbf{High-$T$ region: $T > 100$ K}} \\
\hline
Parameter & $\rho_{0}$ & $a$ & $b$ & -- \\
Units 
& ($\mu\Omega$\,cm) 
& ($\mu\Omega$\,cm\,K$^{-1}$) 
& ($10^{-4}$ $\mu\Omega$\,cm\,K$^{-2}$) 
& -- \\
Value & 135.83 & 0.02 & 1.22 & -- \\
\hline\hline
\end{tabular}}
\end{table}

Generally, in this kind of system, the upturn is contributed by three mechanisms: (1) Weak localization (WL), (2) Electron-electron interaction (EEI), and (3) the Kondo effect. (1) WL is a quantum interference phenomenon that increases back scattering in disordered metals, resulting in an increase in resistivity because of the coherent interference of time-reversed paths as the temperature drops. It shows sensitivity to very minor magnetic fields, which disrupt the interference and diminish the resistivity increase, resulting in negative magnetoresistance. WL scales as $T^{-1/2}$, $-lnT$ and, $-T^{1/2}$ for 1D, 2D and, 3D disordered systems, respectively. (2) EEI refers to the mutual Coulomb interactions that occur between conduction electrons within a material. In an ideal metal characterized by a high carrier density, these interactions are generally shielded and weak. However, in disordered systems, EEI becomes important and can lead to noticeable deviation from standard metallic behavior, particularly at low temperatures which appears as a $T ^{1/2}$ dependence of $\rho$ for three-dimensional (3D) materials. (3) In the Kondo effect, conduction electrons scatter off localized magnetic impurities through spin-flip mechanisms, resulting in a distinct logarithmic rise in resistivity below a specific temperature scale known as the Kondo temperature ($T_K$). The Kondo contribution reaches saturation at low temperatures and is significantly reduced by external magnetic fields, providing a clear experimental indication. Hence, a comprehensive analysis of the low-temperature resistivity measurements and field-dependent transport studies is essential to distinguish among the possible origins of the resistivity upturn. 
\begin{table}[h!]
\caption{Fitting parameters of the low-temperature resistivity ($2<T<75$ K) under 3 T and 9 T magnetic fields.}
\centering
\begin{tabular}{lcccc}
\hline
\hline
\textbf{Field} & $\rho_{0}$ ($\mu\Omega$ cm) & 
$a$ ($10^{-4}\,\mu\Omega$ cm/K$^{2}$) & 
$b$ ($10^{-10}\,\mu\Omega$ cm/K$^{5}$) & 
$c$ ($\mu\Omega$ cm/K$^{1/2}$) \\
\hline
3 T  & 138.11 & 1.67 &1.92  & 0.20 \\
9 T  & 137.70 & 2 & 1.45 & 0.21 \\
\hline
\hline
\end{tabular}
\end{table}

In disordered materials, the Mooij rule indicates that the temperature coefficient of resistivity (TCR) reverses its sign when the resistivity at room temperature surpasses a specific threshold, usually around $\rho_{\mathrm{Th}} \approx 150~\mu\Omega\,\mathrm{cm}$
\cite{Mooij1973}. However, Kaveh and Mott generalized this idea and proposed that there may be a crossover temperature at which the TCR changes sign even at lower temperatures \cite{MKaveh_1982}. The observed increase in resistivity at low temperatures in CoRuFeSi suggests a significant level of disorder, which is consistent with the expectations of the generalized Mooij rule. This phenomenon also indicates that additional scattering mechanisms become active at lower temperatures. 
\clearpage
\subsection{Fermi Surface of the Ordered Structure}

\begin{figure*}[htb]
\centering
\includegraphics[width=0.94\linewidth]{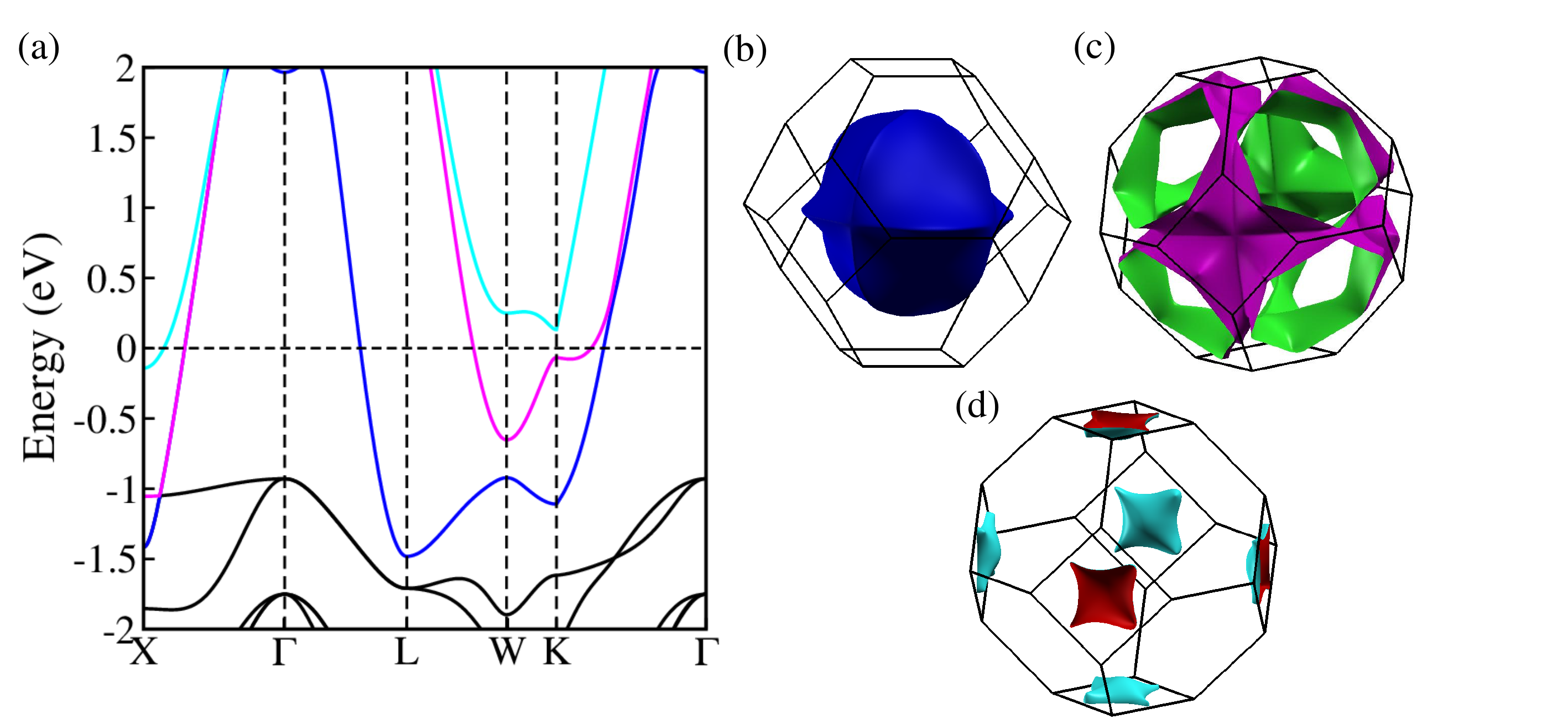}
\caption{(a) Electronic band structure of the spin-up state without SOC, with bands crossing the Fermi level highlighted in color. 
(b-d) Corresponding Fermi surface plots.}
\label{fs}
\end{figure*}

The Fermi surface plot for the three bands crossing the Fermi level ($E_F$) for CoRuFeSi without SOC, shown in  \textcolor{blue}{Fig.~S\ref{fs}(b-d)}, reveals three distinct pockets along the high-symmetry path $X$--$\Gamma$--$L$--$W$--$K$--$\Gamma$, corresponding to three different bands crossing $E_F$.
Among these, the first Fermi surface forms a closed pocket due to band crossings along the $\Gamma$--$X$, $\Gamma$--$L$, and $K$--$\Gamma$ directions, while the second pocket arises from band crossings along the $\Gamma$--$X$, $L$--$W$, and $K$--$\Gamma$ directions. The last pocket exhibits an open character, corresponding to band crossings along the $\Gamma$--$X$ direction. As these bands cross $E_F$ twice, transitioning from the valence band to the conduction band and vice versa, the system exhibits both electron- and hole-like charge carriers, with a dominance of hole-like pockets.

\subsection{ Topological features in ordered structure}

\subsubsection{Nodal line features}
Analogous to the topological features seen in the disordered structure, we analyzed the ordered system, which has not been studied before, and the detailed nontrivial topological aspects are discussed below.

\begin{figure*}[htb]
\centering
\includegraphics[width=0.84\linewidth]{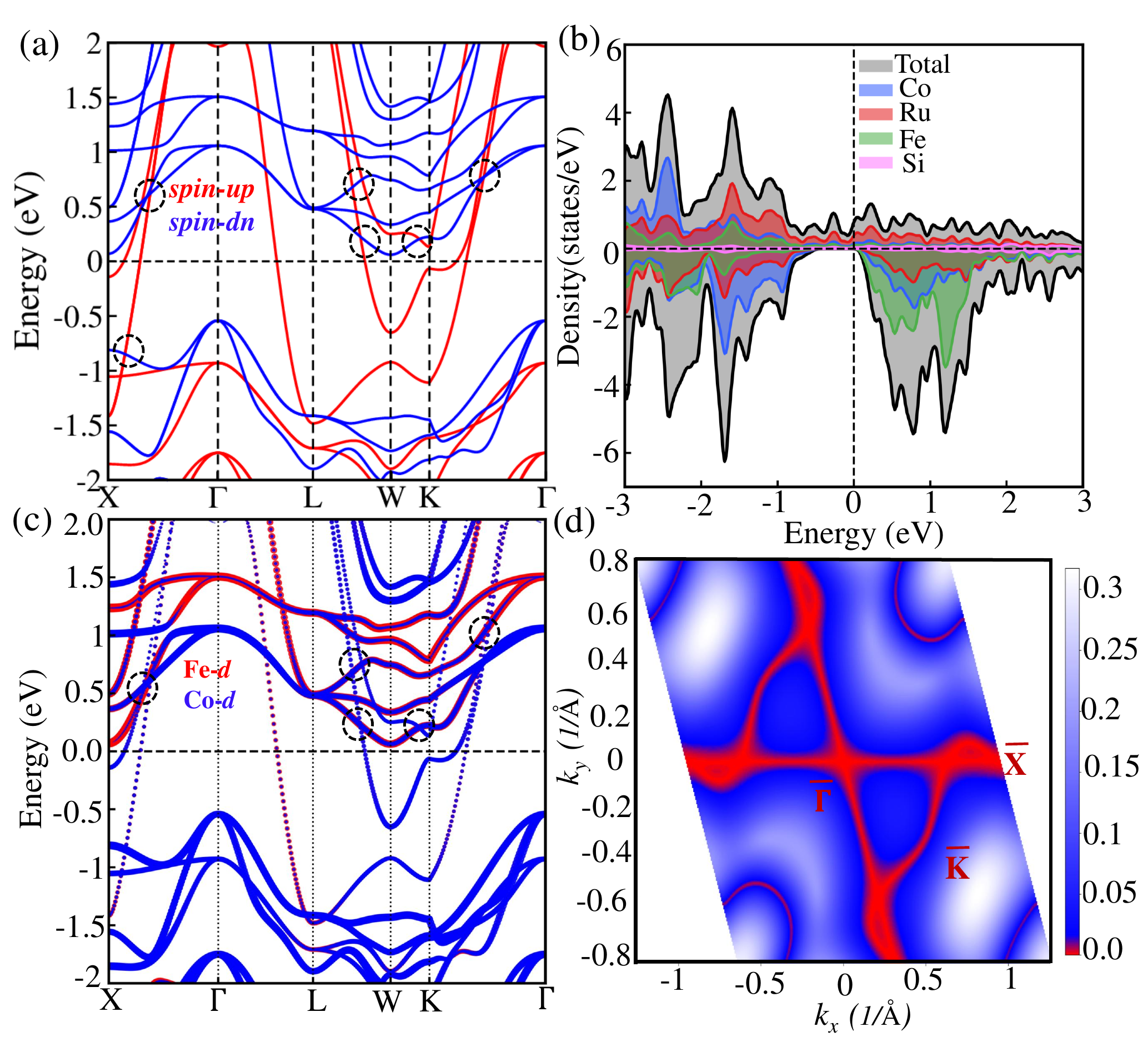}
\caption{(a) Spin-resolved electronic band structure of CoRuFeSi without SOC. (b) Total and projected density of states. (c) Projected band structure showing orbital hybridization in the absence of SOC. (d) Nodal-ring dispersions along the (001) surface.}
\label{nl}
\end{figure*}

The spin-polarized electronic band structure of the CoRuFeSi compound is illustrated in \textcolor{blue}{Fig. S\ref{nl} (a)}, where the spin-up and spin-down channels are shown in red and blue, respectively. The spin-up channel exhibits metallic behavior, whereas the spin-down channel displays semiconducting characteristics, indicating the half-metallic nature of the system. The total density of states (DOS), along with the atom-resolved projected density of states (PDOS), is presented in  \textcolor{blue}{Fig. S\ref{nl} (b)}. Both the valence and conduction bands are primarily contributed by Co, Ru, and Fe atoms, with Co dominating the valence band and Fe the conduction band, while Si contributes negligibly. Furthermore, the reduced DOS for the spin-up channel at the Fermi level reinforces the half-semimetallic  with 100\% spin polarisation character of the system.

The emergence of non-trivial topological features in this system can be attributed to the prominent band crossings near the $\Gamma$--X, L--W, W--K, and $\Gamma$--K high-symmetry paths, as indicated by the dashed circles. These topologically non-trivial features originate from the orbital hybridization between the Co-$d$ and Fe-$d$ states along the $\Gamma$--X path, as shown in the projected band structure in  \textcolor{blue}{Fig. S\ref{nl} (c)}, where the black dashed circles denote states with opposite mirror eigenvalues of $+1$ and $-1$ at the crossing points. Upon inclusion of spin–orbit coupling (SOC), this orbital hybridization induces a band inversion, confirming the presence of nontrivial topological features. These band crossings without SOC may give rise to nodal-line features in the system, which can be further verified through gap-plane calculations.
As shown in  \textcolor{blue}{Fig. S\ref{nl} (d)}, a closed contour with a zero energy gap is observed, revealing nodal-line-like features characterized by a vanishing gap. This nodal line is protected by mirror symmetry. Upon inclusion of spin--orbit coupling (SOC), the mirror symmetries that protect the nodal line are broken, lifting the degeneracy at the crossing points and opening a gap. This gapping of the nodal line gives rise to the emergence of the anomalous Hall effect as discussed in the main manuscript.


\bibliographystyle{unsrt}
\bibliography{CoRuFeSi_supp}